\documentclass{biophys}
\usepackage{helvet,times}
\usepackage{bm,textcomp}
\usepackage{amsfonts,amssymb}
\usepackage[version=3]{mhchem}
\usepackage{xcolor}
\usepackage{setspace}

\jno{kxl014} 
\gridframe{N}
\cropmark{N}

\doi{}

\usepackage{amsmath}

\usepackage[round,numbers,sort&compress]{natbib}

\begin{document}

\setcounter{page}{1} 

\title{Fundamental constraints on the abundances of chemotaxis proteins}

\author{Anne-Florence Bitbol$^{1,2}$ and Ned S. Wingreen$^{1,3}$}

\address{$^1$Lewis-Sigler Institute for Integrative Genomics, $^2$Department of Physics, and $^3$Department of Molecular Biology, Princeton University, Princeton, NJ 08544, USA}



\begin{abstract}%
{Flagellated bacteria, such as \textit{Escherichia coli}, perform directed motion in gradients of concentration of attractants and repellents in a process called chemotaxis. The \textit{E. coli} chemotaxis signaling pathway is a model for signal transduction, but it has unique features. We demonstrate that the need for fast signaling necessitates high abundances of the proteins involved in this pathway. We show that further constraints on the abundances of chemotaxis proteins arise from the requirements of self-assembly, both of flagellar motors and of chemoreceptor arrays. All these constraints are specific to chemotaxis, and published data confirm that chemotaxis proteins tend to be more highly expressed than their homologs in other pathways. Employing a chemotaxis pathway model, we show that the gain of the pathway at the level of the response regulator CheY increases with overall chemotaxis protein abundances. This may explain why, at least in one \textit{E. coli} strain, the abundance of all chemotaxis proteins is higher in media with lower nutrient content. We also demonstrate that the \textit{E. coli} chemotaxis pathway is particularly robust to abundance variations of the motor protein FliM.}
{\today}
{Correspondence: afbitbol@princeton.edu, wingreen@princeton.edu}
\end{abstract}

\maketitle 

\newpage

\section*{Introduction}

Flagellated bacteria such as \textit{Escherichia coli} are able to move up concentration gradients of chemical attractants, and down gradients of repellents, in a process called chemotaxis~\cite{Adler66}. The motion of these bacteria comprises periods of straight swimming called ``runs'', and random changes of direction called ``tumbles''. Run lengths are modulated to yield a three-dimensional random walk biased toward the preferred direction~\cite{Berg72}. Runs occur when flagella rotate counterclockwise and bundle together, while tumbles occur when one or more rotate clockwise and disrupt the bundle~\cite{Turner00}.

In \textit{E. coli}, transmembrane chemoreceptors form large and highly ordered arrays at the cell poles. Chemoreceptors are organized into trimers of dimers, and linked by CheW and CheA into a honeycomb lattice~\cite{Kentner06,Briegel09,Briegel12,Briegel14}, with a 6:1:1 receptor:CheA:CheW stoichiometry in terms of monomers~\cite{Briegel14}. Receptors control the activity of the histidine kinase CheA, which phosphorylates the cytoplasmic response-regulator protein CheY. Phosphorylated CheY (CheY-P) binds to FliM in the flagellar motor to induce clockwise rotation and tumbles. CheA also phosphorylates and activates CheB, a deaminase/methylesterase, that together with the methyltransferase CheR, reversibly modifies specific residues on the receptors to produce adaptation, i.e., to return to a baseline activity level when chemoeffector concentrations stay constant~\cite{Hazelbauer07,Hazelbauer10,Sourjik12}. Upon an increase in the concentration of chemoattractant, the activity of CheA decreases, which leads to fewer tumbles. Conversely, upon a decrease in the concentration of chemoattractant, the activity of CheA increases, yielding more tumbles. This biases the cell's motion toward climbing the gradient of chemoattractant. 

\textit{E. coli} chemotaxis is a model for signal transduction, and is a member of the family of two-component signaling systems that enable bacteria to sense and respond to various features of their environment~\cite{Goulian10,Krell10,Hazelbauer10}. However, the chemotaxis pathway has unique features. First, chemotaxis calls for very fast response times. We demonstrate that this requirement necessitates high abundances of chemotaxis proteins. Second, chemotaxis involves large-scale multi-protein complexes, namely flagellar motors~\cite{Macnab03,Sowa08} and chemoreceptor arrays~\cite{Kentner06,Briegel09,Briegel12,Briegel14}. We show that the consequent self-assembly requirements impose additional constraints on the abundances of chemotaxis proteins. Because of these specific constraints, we hypothesize that chemotaxis proteins will be more highly expressed than their homologs in other pathways. Published data are consistent with this prediction, but more data would be required to definitively confirm it. In addition, using a model of the chemotaxis pathway, we show that the gain of the chemotaxis pathway at the level of CheY-P increases with overall chemotaxis protein abundances. This is consistent with the fact that artificially overexpressing chemotaxis proteins in a concerted manner increases chemotactic efficiency, measured by a swarm assay~\cite{Kollmann05}. Moreover, it may help explain why the abundance of all the chemotaxis proteins can be up to nine-fold higher in nutrient-poor versus rich medium~\cite{Li04}. We also demonstrate that the pathway is particularly robust to abundance variations of the motor protein FliM, in line with other robustness features of the chemotaxis pathway~\cite{Barkai97,Levin98,Alon99,Kollmann05}.

\section*{Models and methods}

\subsubsection*{Chemotaxis pathway model}

We model the \textit{E. coli} chemotaxis signaling pathway by the following system of ordinary differential equations for the average cellular concentrations of each protein in the pathway:
\begin{align}
[\textrm{CheA}]_\textrm{tot}&=[\textrm{CheA}]+[\textrm{CheA-P}]\,, \label{mod1}\\
[\textrm{CheY}]_\textrm{tot}&=[\textrm{CheY}]+[\textrm{CheY-P}]+[\textrm{FliM}\cdot\textrm{CheY-P}]+[\textrm{CheZ}\cdot \textrm{CheY-P}]\,,\\
[\textrm{FliM}]_\textrm{tot}&=[\textrm{FliM}]+[\textrm{FliM}\cdot\textrm{CheY-P}]\,,\\
[\textrm{CheZ}]_\textrm{tot}&=[\textrm{CheZ}]+[\textrm{CheZ}\cdot \textrm{CheY-P}]\,,\\
[\textrm{CheB}]_\textrm{tot}&=[\textrm{CheB}]+[\textrm{CheB-P}]\,,\label{consb}\\
\frac{d[\textrm{CheA-P}]}{dt} &= \alpha k_\textrm{cat}^A [\textrm{CheA}] -  [\textrm{CheA-P}] (k_a^Y [\textrm{CheY}]+ k_a^B [\textrm{CheB}])\,, \label{modalpha}\\
\frac{d[\textrm{CheY-P}]}{dt} &=k_a^Y [\textrm{CheA-P}] [\textrm{CheY}] -[\textrm{CheY-P}]\left(k_a^Z[\textrm{CheZ}]+k_a^M[\textrm{FliM}]+k_h^Y\right)\nonumber\\
& + k_d^Z[\textrm{CheZ}\cdot \textrm{CheY-P}]+ k_d^M [\textrm{FliM}\cdot\textrm{CheY-P}]\,,\\
\frac{d[\textrm{FliM}]}{dt} &= (k_d^M+k_h^Y) [\textrm{FliM}\cdot\textrm{CheY-P}]-k_a^M [\textrm{CheY-P}] [\textrm{FliM}] \,,\label{modflim}\\
\frac{d[\textrm{CheZ}]}{dt} &= (k_d^Z+ k_\textrm{cat}^Z) [\textrm{CheZ}\cdot \textrm{CheY-P}]-k_a^Z [\textrm{CheZ}] [\textrm{CheY-P}] \,,\label{modz}\\
\frac{d[\textrm{CheB-P}]}{dt} &= k_a^B [\textrm{CheA-P}] [\textrm{CheB}] - k_h^B [\textrm{CheB-P}] \label{modf}\,.
\end{align}
Here, concentrations are denoted by square brackets, and total concentrations by ``tot''. Phosphorylated species are denoted by ``-P'', and complexes by a dot between the two species names (e.g., $\textrm{FliM}\cdot\textrm{CheY-P}$). The first five equations express conservation of matter for each protein, while the other ones convey the kinetics of the chemical reactions in the pathway. These reactions are depicted in Eqs.~\ref{eqch1}-\ref{eqchf} of the Supporting Material. 

We focus on the adapted state of the pathway and on its initial response to attractant or repellent, without explicitly modeling the slower dynamics of adaptation. In the adapted state, the active fraction $\alpha$ of CheA is modeled as~\cite{Meir10}:
\begin{equation}
 \alpha=\frac{k_\textrm{cat}^R[\textrm{CheR}]_\textrm{tot}}{k_\textrm{cat}^R[\textrm{CheR}]_\textrm{tot}+k_\textrm{cat}^B[\textrm{CheB-P}]}\,,\label{alpha}
\end{equation}
which follows if CheR methylates inactive receptors and CheB-P demethylates active receptors. This active fraction is taken into account in the system of differential equations in Eqs.~\ref{mod1}-\ref{modf} through the reduction of the time-averaged autocatalytic rate of CheA from $k_\textrm{cat}^A$ to $\alpha k_\textrm{cat}^A$ (see Eq.~\ref{modalpha}), as in Ref.~\cite{Sourjik02}. 

\paragraph*{Parameter values.}
We use experimentally-determined values for the reaction rates $k$ in Eqs.~\ref{mod1}-\ref{alpha}, except for $k_a^Z$ and $k_\textrm{cat}^Z$ (Table~\ref{KinValues}). Indeed, while the reaction rates for CheY-P dephosphorylation by CheZ have been measured \textit{in vitro} in the absence of CheA~\cite{Silversmith08}, it is known that CheZ binds to CheA-short, a translational variant of CheA that cannot autophosphorylate, and that this binding significantly activates CheZ~\cite{Wang96,OConnor04,Vaknin04}. We thus adjusted $k_a^Z$ and $k_\textrm{cat}^Z$ in order to obtain a fraction of CheZ bound to CheY-P of $\sim 30\%$, consistent with \textit{in vivo} FRET measurements in the adapted steady state~\cite{Sourjik02,Shimizu10}.

\begin{figure}[h t b]
\centerline{\includegraphics[width=0.45\textwidth]{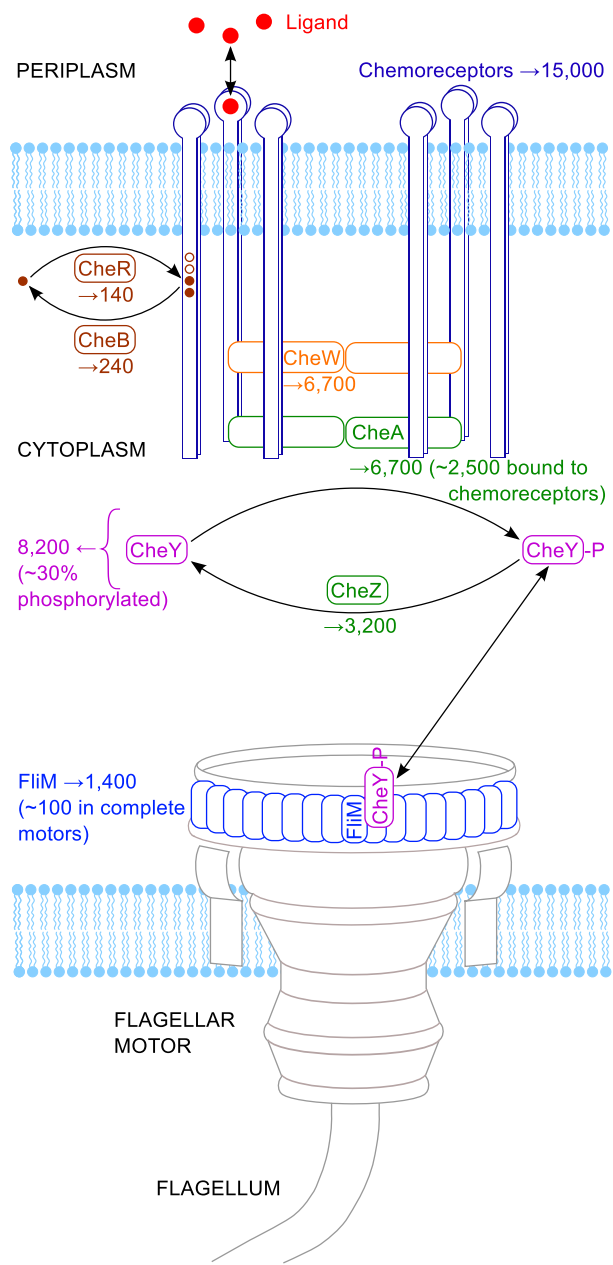}}
\caption{\label{Fig1}Schematic of the chemotaxis signaling pathway in \textit{E. coli}. The number of copies per cell is indicated for each protein in the pathway. These numbers correspond to the measurements on strain RP437 in rich medium in Ref.~\cite{Li04} for all proteins but FliM, and to the measurements in Refs.~\cite{Tang95,Delalez10} for FliM, also in rich medium.}
\end{figure}

We use the average copy numbers of each chemotaxis protein per cell measured in Ref.~\cite{Li04} for strain RP437 in rich medium for all proteins but FliM, and those in Refs.~\cite{Tang95,Delalez10} for FliM, also in rich medium (Fig.~\ref{Fig1}). Importantly, the autocatalytic rate $k_\mathrm{cat}^A$ of the histidine kinase CheA is increased about 100-fold when CheA is in complex with chemoreceptors and CheW~\cite{Levit99}, so only CheA in signaling complexes has significant kinase activity. Receptors are limiting~\cite{Li04,Endres08} for signaling complexes with a 6:1:1 receptor:CheA:CheW stoichiometry~\cite{Briegel14}. Hence, we consider that the total number of CheA proteins per cell that can be active (setting $[\textrm{CheA}]_\textrm{tot}$) is one-sixth the total number of receptor monomers. It is also observed that less than 30\% of FliM is found in complete flagellar motors~\cite{Tang95,Zhao96,Delalez10}, and that only 16\% is in the soluble fraction~\cite{Zhao96}, while more than 25\% of FliM~\cite{Zhao96,Delalez10}, probably all the rest, is found in partially assembled structures (see Supporting Material). Isolated FliM molecules have a much lower affinity for CheY-P than FliM in motors, with a dissociation constant of 27~$\mu\textrm{M}$~\cite{McEvoy99} versus 3.5~$\mu\textrm{M}$~\cite{Cluzel00,Sourjik02,Sagi03,Yuan13}, which leads us to disregard isolated FliM. In the absence of any data to the contrary, we assume that CheY-P binds FliM in partly and fully-assembled motors with the same affinity. For each chemotaxis pathway protein, we derive the corresponding effective total cellular concentration using the standard \textit{E. coli} cell volume of 1.4~fL~\cite{Sourjik02,Kollmann05} (Table~\ref{Conc}). 

\paragraph*{Numerical solution.}
We solve Eqs.~\ref{mod1}-\ref{alpha} at steady state numerically using `NSolve' (Wolfram Mathematica). The initial response to saturating attractant (or repellent) is obtained by abruptly decreasing the CheA active fraction $\alpha$ to 0 (or increasing it to 1) from its adapted value. Hence, we solve Eqs.~\ref{mod1}-\ref{modf} numerically with $\alpha=0$ (or 1), with the adapted concentrations as initial conditions, using `NDSolve' (Wolfram Mathematica).

\paragraph*{Pathway gain.}
We are interested in the gain of the chemotaxis pathway. The input is the active fraction $\alpha$ of CheA, which directly depends on receptor states and hence on chemoeffector concentrations. We consider two different outputs: the concentration $[\textrm{CheY-P}]$ of phosphorylated CheY, and the fraction $\psi$ of FliM molecules bound to CheY-P, with corresponding gains defined by 
\begin{equation}
G_\textrm{CheY-P}=\frac{\Delta[\textrm{CheY-P}]/[\textrm{CheY-P}]}{\Delta\alpha/\alpha}\,, \label{GYP}
\end{equation}	
and	
\begin{equation}
G_\psi=\frac{\Delta\psi/\psi}{\Delta\alpha/\alpha}\,. \label{Gpsi}
\end{equation}

In practice, gains in the linear-response regime are computed for the pre-adaptation response to a 1\% increase of the CheA active fraction $\alpha$ from its adapted value determined by Eq.~\ref{alpha}.

\clearpage

\section*{Results}

\subsubsection*{Fast response imposes constraints on the abundances of chemotaxis proteins}

\begin{figure}[h t b]
\centering
\includegraphics[width=0.4\textwidth]{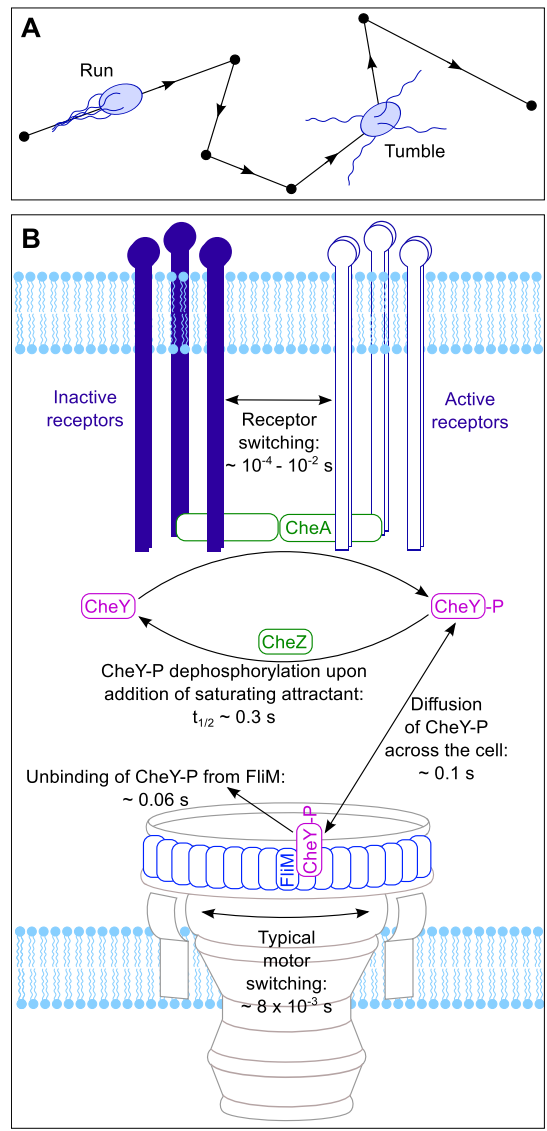}
\caption{\label{Fig2}Response timescales in the chemotactic pathway. \textbf{A.} Schematic of a chemotactic trajectory: the bacterium swims straight during ``runs'' (lines with arrows), and randomly changes direction during ``tumbles'' (dots), resulting in a three-dimensional random walk. The mean run time under adapted conditions is about 1 s~\cite{Alon98}. \textbf{B.} Schematic of the timescales involved in the initial (pre-adaptation) response to saturating attractant. The longest timescale corresponds to CheY-P dephosphorylation by CheZ~\cite{Sourjik02}: it is much longer than receptor switching~\cite{Skoge11,Lan12}, motor switching~\cite{Bai10}, and unbinding of CheY-P from FliM (see main text), and slightly longer than CheY-P diffusion~\cite{Elowitz99,Nenninger10}.}
\end{figure}

Chemotactic trajectories are composed of straight ``runs'' and  random changes of directions or ``tumbles'' (Fig.~\ref{Fig2}A). The mean run time of \textit{E. coli} cells under adapted conditions is about one second~\cite{Alon98}. Hence, in practice, cells must make a decision whether to change direction in less than a second. The observed timescale of response to a saturating attractant is $\sim0.3$~seconds~\cite{Sourjik02}. Fig.~\ref{Fig2}B shows the timescales of the different molecular events involved in this response. The longest one is the dephosphorylation time $\sim0.3\,\textrm{s}$ of the cellular pool of CheY-P by the phosphatase CheZ~\cite{Sourjik02}. Here, we show that this timescale implies lower bounds on the dissociation constant of FliM and CheY-P and on the abundances of several proteins in the chemotaxis pathway. 

The CheY-P molecules bound to FliM proteins need to unbind and to be dephosphorylated within this 0.3~s for the pool of CheY-P to reflect the current chemoeffector concentration, thus ensuring an appropriate response. The unbinding timescale is $1/k_d^M$, where $k_d^M$ is the dissociation rate of FliM and CheY-P (see Eq.~\ref{modflim}), so $1/k_d^M\lesssim0.3~\textrm{s}$ implies $k_d^M\gtrsim 3.3~/\textrm{s}$. Since the binding of FliM and CheY-P is diffusion-limited, i.e. as fast as it can be, with a rate constant $k_a^M=5~/\textrm{s}/\mu\textrm{M}$~\cite{Sourjik02, Northrup92}, the dissociation constant of FliM and CheY-P must satisfy $K_d^M\equiv k_d^M/k_a^M\gtrsim 0.7~\mu\textrm{M}$. In reality, $K_d^M=3.5~\mu\textrm{M}$~\cite{Cluzel00,Sourjik02,Sagi03,Yuan13}, and the associated unbinding timescale is 0.06~s. Hence, our lower bound on $K_d^M$ is satisfied.

In the adapted state, the fraction $\psi$ of FliM molecules that are bound to CheY-P should be in the intermediate range, in order to respond readily to both increases and decreases of the free CheY-P concentration  $[\textrm{CheY-P}]$. Assuming an adapted $\psi\gtrsim 0.25$, which is in the lower range of the region where the motor can switch rotation direction~\cite{Sourjik02}, we obtain $[\textrm{CheY-P}]\gtrsim K_d^M/3=1.17~\mu\textrm{M}$, and $[\textrm{FliM}\cdot\textrm{CheY-P}]\gtrsim 0.25\, [\textrm{FliM}]_\textrm{tot}= 0.35~\mu\textrm{M}$, where we used the total FliM concentration in Table~\ref{Conc} (see Methods and Models). Hence, the total cellular concentration of CheY-P is $C_\textrm{CheY-P}=[\textrm{CheY-P}]+[\textrm{FliM}\cdot\textrm{CheY-P}]\gtrsim1.5~\mu\textrm{M}$. Note that here we do not take into account the CheY-P that are bound to CheZ and thus essentially sure to be dephosphorylated (since $k_\mathrm{cat}^Z\gg k_d^Z$, see Table~\ref{KinValues}). In practice, about 30\% of CheY is phosphorylated~\cite{Alon98}, yielding $C_\textrm{CheY-P}=3~\mu\textrm{M}$ (using the total CheY concentration in Table~\ref{Conc}). Hence, our lower bound on $C_\textrm{CheY-P}$ is satisfied, with the actual value being only twice as large.

We now focus on the dephosphorylation of CheY-P, whose steady-state rate is (Eq.~\ref{modz})
\begin{equation}
\left.\frac{d[\textrm{CheY-P}]}{dt}\right|_\textrm{dephos}=-k_\mathrm{cat}^Z [\textrm{CheZ}]_\textrm{tot} \,\frac{[\textrm{CheY-P}]}{\frac{k_\mathrm{cat}^Z+k_d^Z}{k_a^Z}+[\textrm{CheY-P}]}\,. \label{vdephos}
\end{equation}
The whole cellular pool of non-CheZ-bound CheY-P, with concentration $C_\textrm{CheY-P}$, needs to be dephosphorylated within 0.3~s. Using the minimal values of $[\textrm{CheY-P}]$ and $C_\textrm{CheY-P}$ calculated above and rate constants in Table~\ref{KinValues}, this requirement yields $ [\textrm{CheZ}]_\textrm{tot}\gtrsim 2.3~\mu\textrm{M}$. Note that using the experimental values for $k_\mathrm{cat}^Z$ and $k_a^Z$ from Ref.~\cite{Silversmith08}, which disregard CheZ activation by CheA-short, gives a similar result: $ [\textrm{CheZ}]_\textrm{tot}\gtrsim 1.8~\mu\textrm{M}$. Experiments yield $[\textrm{CheZ}]_\textrm{tot}=3.8~\mu\textrm{M}$ (Table~\ref{Conc}), so here too, our lower bound is satisfied, with the actual value being less than twice as large.

For turnover to occur within 0.3~s, ensuring that $[\textrm{CheY-P}]$ reflects the current chemoeffector concentration, the whole cellular pool of non-CheZ-bound CheY-P also needs to be (re)phosphorylated within this time. Phosphotransfer from CheA-P to CheY being very fast, the limiting step is CheA autophosphorylation~\cite{Mayover99}. Hence, the steady-state CheY phosphorylation rate is simply $\alpha k_\textrm{cat}^A[\textrm{CheA}]_\textrm{tot}$ (Eq.~\ref{modalpha}). Using the minimal values of $C_\textrm{CheY-P}$ and of $ [\textrm{CheZ}]_\textrm{tot}$ calculated above, and $k_\textrm{cat}^A=20\,/\textrm{s}$ (Table~\ref{KinValues}), we obtain $\alpha[\textrm{CheA}]_\textrm{tot}\gtrsim 0.25~\mu\textrm{M}$. The total concentration of CheA in arrays (determined from the receptor concentration and the stoichiometry, see Models and methods and Table~\ref{Conc}), $[\textrm{CheA}]_\textrm{tot}=3.0~\mu\textrm{M}$, is substantially larger than this lower bound. This hints at a low active fraction $\alpha$, consistent with previous estimates, which range from a few percent~\cite{Sourjik02,Neumann14} to about 30\%~\cite{Sourjik02b,Shimizu10}.

Hence, the requirements of fast signaling impose lower bounds on the cellular concentrations of CheY-P, CheZ, and active CheA, as well as on the dissociation constant $K_d^M$ of CheY-P and FliM. These lower bounds are satisfied by experimental values, and are consistent with a low adapted CheA active fraction.

\subsubsection*{Pathway model accounts for observed concentrations and response times}
While the above simple arguments enabled us to derive constraints on the abundances of chemotaxis proteins, a more detailed comparison to observed concentrations and response times requires a mathematical model. Here we present results from the pathway model given by Eqs.~\ref{mod1}-\ref{alpha}. Similar models have been productively employed previously to investigate various aspects of the chemotaxis network~\cite{Levin98,Sourjik02,Kollmann05,Oleksiuk11}. Our focus is on the impact of protein abundances on gain.

The adapted steady-state of the chemotaxis pathway is obtained by solving Eqs.~\ref{mod1}-\ref{alpha} at steady-state with the parameter values in Tables~\ref{KinValues} and~\ref{Conc} (see Models and methods). It yields $[\textrm{CheY-P}]+[\textrm{FliM}\cdot\textrm{CheY-P}]=3.0\,\mu \textrm{M}$, in agreement with Ref.~\cite{Cluzel00}, and a proportion of phosphorylated CheY-P of 31\%, in agreement with Ref.~\cite{Alon98}. Besides, we obtain a fraction $\psi$ of FliM molecules that are bound to CheY-P of 41\% in the adapted state, which is in the functional range where the flagellar motor is able to switch~\cite{Sourjik02}. We also obtain a fraction $\alpha$ of active CheA  of 25\% in the adapted state, within the range of previous estimates~\cite{Sourjik02,Neumann14,Sourjik02b,Shimizu10}.

\begin{figure}[h t b]
\centering
\includegraphics[width=0.45\textwidth]{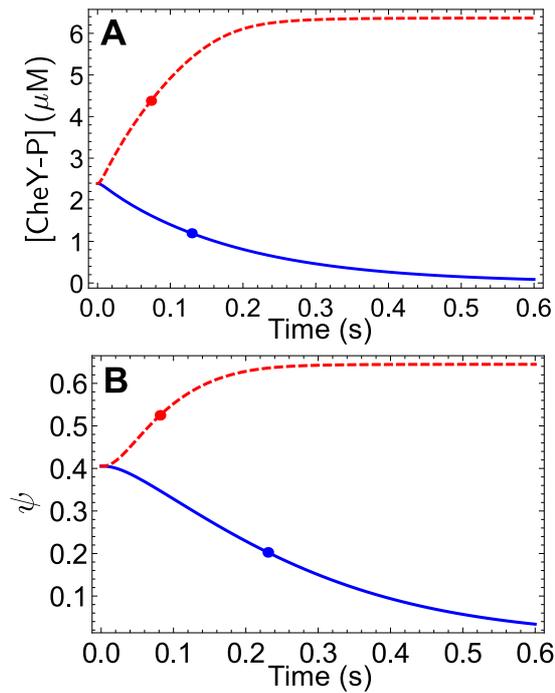}
\caption{\label{Fig3}Response to addition of saturating attractant or repellent, obtained from the pathway model (Eqs.~\ref{mod1}-\ref{modf}). \textbf{A.} Concentration of free CheY-P ($[\textrm{CheY-P}]$) versus time after a step addition of saturating attractant (blue curve) or repellent (red dashed curve). Addition of saturating attractant (repellent) is modeled by changing instantaneously the CheA active fraction, $\alpha$, from its adapted value (cf. Eq.~\ref{alpha}) to 0 (1). Dots indicate half-maximal response. \textbf{B.} Fraction $\psi$ of FliM proteins bound to CheY-P versus time after a step addition of saturating attractant (blue curve) or repellent (red dashed curve). Dots indicate half-maximal response. }
\end{figure}

The initial (pre-adaptation) response of the pathway to instantaneous addition of saturating attractant (or repellent) is obtained by solving Eqs.~\ref{mod1}-\ref{modf} with the adapted steady-state concentrations as initial conditions, setting the CheA active fraction $\alpha$ to 0 (or 1) (see Models and methods). Upon addition of attractant, $[\textrm{CheY-P}]$ is found to decrease to 0 with a half-time of 0.13~s, and the fraction $\psi$ of FliM proteins bound to CheY-P decreases with a half-time of 0.23~s (Fig.~\ref{Fig3}). This is in reasonable agreement with Ref.~\cite{Sourjik02}, where the half-time for the decay of CheY-P bound to FliM, observed experimentally by FRET, is 0.32~s. Note that the difference between the timescales obtained for $[\textrm{CheY-P}]$ and for $\psi$ from our pathway model indicates that the unbinding time of CheY-P from FliM is not negligible, contrary to the usual assumption~\cite{Sourjik02}. Addition of repellent yields a faster response, with half-times of 0.07~s and 0.08~s for the respective increases of $[\textrm{CheY-P}]$ and of $\psi$ (Fig.~\ref{Fig3}). This is in reasonable agreement with the experimental value of 0.03~s for the half-time of the increase of $\psi$~\cite{Sourjik02}. Response to saturating repellent is faster because it relies on CheY phosphorylation by CheA, which is very fast when $\alpha=1$, while CheY-P dephosphorylation is limiting in response to attractant (Fig.~\ref{Fig2}).

The good agreement of the model with observations, obtained by adjusting only $k_\textrm{cat}^Z$ and $k_a^Z$ to match the fraction of CheZ bound to CheY-P (see Models and methods), encourages us to further study the model's implications.

\newpage

\subsubsection*{Effect of a concerted increase of protein abundances}

\begin{figure}[h t b]
\centering
\includegraphics[width=0.45\textwidth]{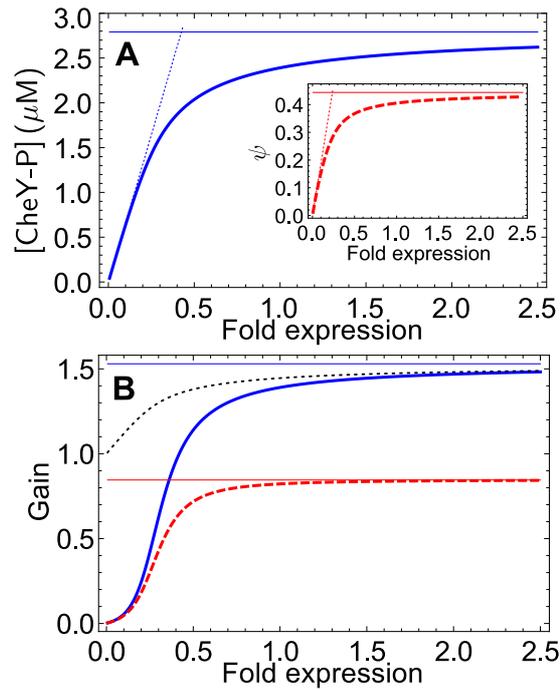}
\caption{\label{Fig4}Effects of fold-change of expression of all chemotaxis signaling proteins (as in Ref.~\cite{Kollmann05}), obtained from the pathway model in the adapted state (Eqs.~\ref{mod1}-\ref{alpha}). One-fold expression corresponds to the abundances in Table~\ref{Conc}, i.e. to those measured in Ref.~\cite{Li04} for strain RP437 in rich medium. In the same rich medium, the fold expression for strain OW1 is about 0.22, while in minimal medium, the fold expression is about 1.1 for strain RP437 and 2.0 for strain OW1~\cite{Li04} (values averaged over all chemotaxis signaling proteins). \textbf{A.} Adapted free CheY-P concentration ($[\textrm{CheY-P}]$) versus fold expression of the chemotaxis proteins. Inset: adapted fraction $\psi$ of FliM proteins bound to CheY-P versus fold expression. Thin horizontal lines: analytical high-abundance limit in the fast-phosphotransfer regime (Eqs.~\ref{ypFPmt} and~\ref{psiFP}). Thin dotted lines: analytical low-abundance limit (Eqs.~\ref{sollowmt} and~\ref{sollowb}).
\textbf{B.}  Corresponding gain in the linear-response regime. Blue curve: gain for CheY-P, $G_\textrm{CheY-P}$ (Eq.~\ref{GYP}). Red (dashed) curve: gain for $\psi$, $G_\psi$ (Eq.~\ref{Gpsi}). Thin horizontal lines: analytical high-abundance asymptotic gains in the fast-phosphotransfer regime (Eqs.~\ref{GS1mt} and~\ref{GS2}). Dotted curve: ratio of total CheZ concentration to free CheZ concentration, $[\textrm{CheZ}]_\textrm{tot}/[\textrm{CheZ}]$; in the simplified-pathway fast-phosphotransfer regime, $G_\textrm{CheY-P}=[\textrm{CheZ}]_\textrm{tot}/[\textrm{CheZ}]$ (see Supporting Material).}
\end{figure}

The overall abundances of chemotaxis signaling proteins (Che proteins and chemoreceptors) are variable across \textit{E. coli} strains and growth conditions, but relative proportions are well-conserved~\cite{Li04}. Strikingly, these proteins are more highly expressed in minimal medium than in rich medium~\cite{Li04}. When the abundances of chemotaxis signaling proteins were varied in a concerted fashion~\cite{Kollmann05}, the chemotactic efficiency of cells (measured by a swarm assay) was found to increase sharply up to wild-type abundance, and then to continue increasing much more gradually while progressively leveling off. Here, to mimic the experiment of Ref.~\cite{Kollmann05}, we vary the abundances of CheA, CheY, CheZ, CheB and CheR, while keeping their proportions and the FliM abundance fixed, as in Table~\ref{Conc}. (We checked that varying the abundance of FliM in a concerted fashion with the rest does not affect our conclusions.) Solving our pathway model Eqs.~\ref{mod1}-\ref{alpha} in the adapted steady state, we find that when protein abundances are increased, $[\textrm{CheY-P}]$ and $\psi$ both increase sharply up to about reference abundances, and the increase then progressively levels off (Fig.~\ref{Fig4}A). Our reference abundances (one-fold in Fig.~\ref{Fig4}) correspond to those measured in Ref.~\cite{Li04} for strain RP437 in rich medium (see Models and methods and Table~\ref{Conc}). 

The effect of a concerted variation of protein abundances on $[\textrm{CheY-P}]$ was previously modeled in Ref.~\cite{Levin98}. Our results (Fig.~\ref{Fig4}A) are mostly consistent with Ref.~\cite{Levin98}, but using one adaptation model Ref.~\cite{Levin98} obtained a maximum in $[\textrm{CheY-P}]$ versus fold abundance. In our framework too, modifying details of the adaptation model (Eq.~\ref{alpha}) can result in such a maximum, but above one-fold expression (for realistic parameter values), so our main conclusions are not affected. In Ref.~\cite{Kollmann05}, clockwise bias was found to be monotonic versus concerted fold expression, which is consistent with our results (Fig.~\ref{Fig4}A). Building on a similar framework to Ref.~\cite{Levin98}, we include CheZ saturation by CheY-P, which has now been measured~\cite{Sourjik02,Shimizu10}, and we discuss FliM occupancy $\psi$ and gain, and provide analytical insight for simple regimes. 

For reference abundances and higher, the steady-state phosphorylated fraction of CheA is very small, because of the rapidity of phosphotransfer from CheA-P to CheY~\cite{Mayover99}. In this ``fast phosphotransfer regime'', it is possible to solve analytically a simplified version of the pathway (see Supporting Material): if the auto-phosphorylation rate of CheA is less than the maximal dephosphorylation rate of CheY-P by CheZ, i.e. if
\begin{equation}
\alpha<\frac{k_\textrm{cat}^Z[\textrm{CheZ}]_\textrm{tot}}{k_\textrm{cat}^A[\textrm{CheA}]_\textrm{tot}}\,,\label{cdnFPmt}
\end{equation}
then
\begin{equation}
[\textrm{CheY-P}]=\frac{\alpha \frac{k_\textrm{cat}^A}{k_a^Z}}{\frac{[\textrm{CheZ}]_\textrm{tot}}{[\textrm{CheA}]_\textrm{tot}}-\alpha\frac{k_\textrm{cat}^A}{k_\textrm{cat}^Z}}\,,\label{ypFPmt}
\end{equation}
and $\psi$ is given by Eq.~\ref{psiFP}. These expressions only depend on abundance ratios, on kinetic rate constants, and on $\alpha$, which converges to a constant value at high abundances (see Supporting Material). Hence, in the high-abundance limit, these steady-state values of $[\textrm{CheY-P}]$ and $\psi$, which arise from the equilibration of phosphorylation and dephosphorylation of CheY, converge to plateaus invariant to concerted variations of the overall abundances. Conversely, if the condition in Eq.~\ref{cdnFPmt} is violated, CheZ is saturated, and $[\textrm{CheY-P}]$ increases with overall abundances. The conditions for the fast phosphotransfer regime are satisfied with the standard abundances used here and with higher overall abundances (see Supporting Material). The plateaus of $[\textrm{CheY-P}]$ (Eq.~\ref{ypFPmt}) and $\psi$ (Eq.~\ref{psiFP}) are indicated by thin lines in Fig.~\ref{Fig4}A. 

In the opposite limit of low abundances, two-molecule encounters become unlikely, including the binding of CheY to CheA-P, so the phosphorylated fraction of CheA becomes high, and only a small fraction of CheZ and of FliM are bound to CheY-P. Using the simplified pathway model presented in the Supporting Material, we show that if $[\textrm{CheY}]_\textrm{tot}\ll \min(\alpha k_\textrm{cat}^A/k_a^Y,\,k_\textrm{cat}^Z/k_a^Z,\,K_d^M)$ and $[\textrm{CheZ}]_\textrm{tot}\ll k_\textrm{cat}^Z/k_a^Z$, then
\begin{equation}
[\textrm{CheY-P}]=\frac{[\textrm{CheY}]_\textrm{tot}}{1+\frac{k_a^M}{k_d^M}[\textrm{FliM}]_\textrm{tot}+\frac{k_a^Z}{k_a^Y}\frac{[\textrm{CheZ}]_\textrm{tot}}{[\textrm{CheA}]_\textrm{tot}}}\,. \label{sollowmt}
\end{equation}
Hence, in the low-abundance limit, $[\textrm{CheY-P}]$ grows in proportion with the overall abundances of the Che proteins. The same is true for $\psi$ (Eq.~\ref{sollowb}). The low-abundance asymptotes Eqs.~\ref{sollowmt} and~\ref{sollowb} are plotted as thin dotted lines in Fig.~\ref{Fig4}A.

Our pathway model also yields the gain $G_\textrm{CheY-P}$ of the pathway at the level of the response regulator. This gain grows with overall abundance of chemotaxis proteins, and plateaus in the high-abundance limit (Fig.~\ref{Fig4}B). The corresponding asymptotic value can be determined analytically within the simplified pathway model in the fast-phosphotransfer regime: $G_\textrm{CheY-P}$ (Eq.~\ref{GYP}) can be obtained from Eq.~\ref{ypFPmt}. It yields
\begin{equation}
G_\textrm{CheY-P}=\frac{\frac{[\textrm{CheZ}]_\textrm{tot}}{[\textrm{CheA}]_\textrm{tot}}}{\frac{[\textrm{CheZ}]_\textrm{tot}}{[\textrm{CheA}]_\textrm{tot}}-\alpha\frac{k_\textrm{cat}^A}{k_\textrm{cat}^Z}}\,,\label{GS1mt}
\end{equation}
which becomes independent of overall abundances as $\alpha$ converges to its high-abundance limit. Besides, in this regime, it can be shown that $G_\textrm{CheY-P}=[\textrm{CheZ}]_\textrm{tot}/[\textrm{CheZ}]$ (see Supporting Material). Thus, the gain in $[\textrm{CheY-P}]$ arises from the saturation of the phosphatase CheZ by CheY-P: increasing the active fraction $\alpha$ of CheA increases phosphotransfer to CheY, and hence $[\textrm{CheY-P}]$, but this increase is larger than that of $\alpha$ because, at the same time, CheZ becomes more saturated, reducing the rate of dephosphorylation of CheY-P (see also Ref.~\cite{vanAlbada09}). In Fig.~\ref{Fig4}B, the thin horizontal blue line represents the plateau for $G_\textrm{CheY-P}$ (Eq.~\ref{GS1mt}), and the dotted curve shows $[\textrm{CheZ}]_\textrm{tot}/[\textrm{CheZ}]$: at sufficiently high abundances, it closely approximates the gain derived from numerical solution of the full pathway. Similarly, $G_\psi$ (Eq.~\ref{Gpsi}) can be determined analytically within the simplified pathway model in the fast-phosphotransfer regime (Eq.~\ref{GS2}, thin horizontal red line in Fig.~\ref{Fig4}B).

We conclude that the gain in CheY-P increases with overall abundances, up to about reference levels. Moreover, chemotactic signaling is robust with respect to concerted overexpression of the chemotaxis proteins (see also Ref.~\cite{Kollmann05}), as $[\textrm{CheY-P}]$ remains lower than $K_d^M=3.5~\mu\textrm{M}$, so that $\psi<0.5$ remains in the functional range, below the threshold value (about 0.57) above which the motor only rotates clockwise~\cite{Sourjik02}.

\subsubsection*{Effect of separately varying the concentration of each protein in the pathway}

\begin{figure}[h t b]
\centering
\includegraphics[width=0.45\textwidth]{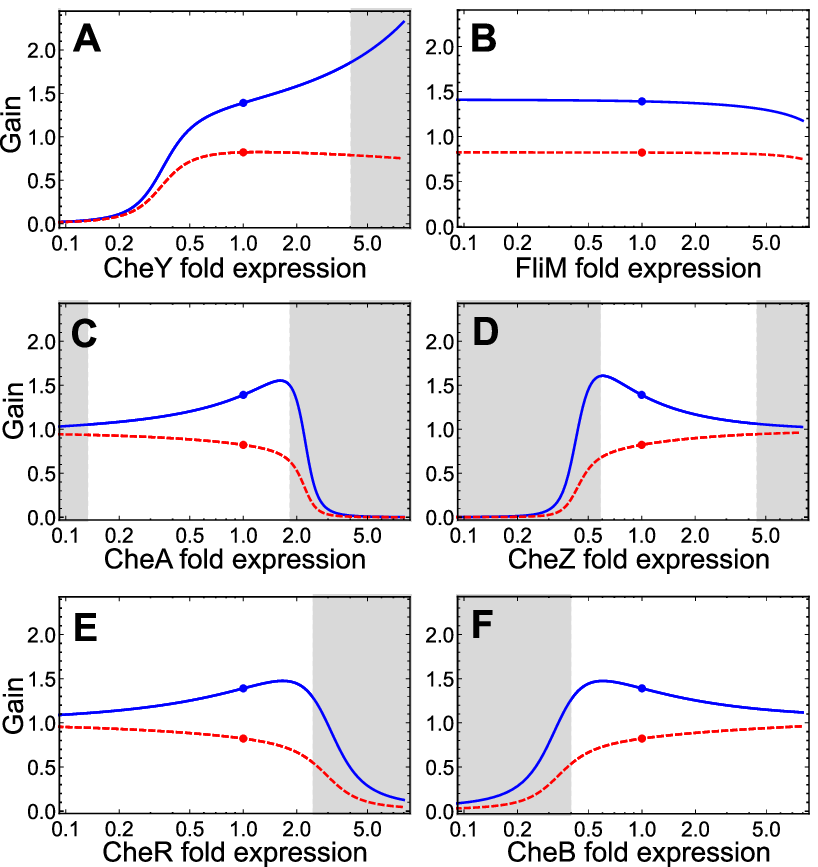}
\caption{\label{Fig5}Effect of fold-change of expression of each chemotaxis signaling protein separately, obtained from the pathway model in the adapted state (Eqs.~\ref{mod1}-\ref{alpha}). One-fold expression corresponds to the abundances in Table~\ref{Conc}, i.e. to those measured in Ref.~\cite{Li04} for strain RP437 in rich medium, as in Fig.~\ref{Fig4}. \textbf{A}-\textbf{F} (linear-log plots).  Blue curves: gain for CheY-P, $G_\textrm{CheY-P}$ (Eq.~\ref{GYP}), versus fold expression of each protein, keeping all others at their one-fold level. Red dashed curves: gain for $\psi$, $G_\psi$ (Eq.~\ref{Gpsi}).  In the shaded zones, $\psi$ is either smaller than 0.11 or larger than 0.57 (Fig.~\ref{FigS2}), in which case the flagellar motor should rotate only counterclockwise or only clockwise, respectively, in the adapted state~\cite{Sourjik02}.}
\end{figure}

To study the effect of varying the abundance of each protein separately on the adapted steady state of the pathway, we separately varied CheY, FliM, CheA, CheZ, CheR, or CheB abundances, while keeping the abundances of all others fixed (values in Table~\ref{Conc}). Specifically, we calculated the gains $G_\textrm{CheY-P}$ and $G_\psi$ (Fig.~\ref{Fig5}), as well as the fraction $\psi$ of FliM molecules bound to CheY-P (Fig.~\ref{FigS2}) and the concentration $[\textrm{CheY-P}]$ of free CheY-P (Fig.~\ref{FigS2_yp}). 

The effect on $[\textrm{CheY-P}]$ of protein abundance variations was investigated in Ref.~\cite{Levin98}. In addition to the differences mentioned above, this previous study did not include FliM, but included CheW and receptors. Our results (Fig.~\ref{FigS2_yp}) are consistent with those of Ref.~\cite{Levin98} for abundance variations of CheY, CheZ, CheR, and CheB. However, Ref.~\cite{Levin98} obtained a weak maximum of $[\textrm{CheY-P}]$ upon CheA abundance variation, arising from their model of CheA interactions with CheW and receptors. We focus on gain, and on the stability of the pathway to FliM abundance variation, which were not included in Ref.~\cite{Levin98}.

Fig.~\ref{Fig5} shows that the gain of the chemotaxis pathway is robust to moderate individual variations of the abundances of each protein. Variations of $G_\psi$ are even weaker than those of $G_\textrm{CheY-P}$, due to the mitigating effect of FliM saturation by CheY-P.

Fig.~\ref{Fig5}A shows that $G_\textrm{CheY-P}$ increases with CheY abundance. Indeed, increased CheY abundance (at constant total CheB and CheA levels) results in less CheB phosphorylation, due to competition for CheA-P. Reduced $[\textrm{CheB-P}]$ in turn results in an increase of the adapted $\alpha$ (Eq.~\ref{alpha}), and hence of $[\textrm{CheY-P}]$ and $\psi$ (Fig.~\ref{FigS2}A). Higher $[\textrm{CheY-P}]$ (at constant total CheZ level) means that CheZ becomes more saturated, increasing $G_\textrm{CheY-P}$ (see above). Consistently, in the fast phosphotransfer regime, Eq.~\ref{GS1mt} shows that $G_\textrm{CheY-P}$ is an increasing function of $\alpha$, which itself increases with CheY abundance, for the above-mentioned reasons. 

Fig.~\ref{Fig5}B shows that the gains are almost independent of the abundance of FliM, and Fig.~\ref{FigS2}B shows that the same is true for $\psi$. In addition, solving our simplified pathway model in the fast phosphotransfer regime gives expressions for $[\textrm{CheY-P}]$, for $\psi$, and for the gains that are entirely independent of FliM abundances (Eqs.~\ref{ypFPmt} and~\ref{psiFP},~\ref{GS1mt} and~\ref{GS2}). This robustness of the pathway to FliM abundance variation arises from the fact that, in contrast to the free CheY-P molecules, the ones that are bound to FliM cannot be dephosphorylated by CheZ (they can auto-dephosphorylate, but this process is much slower than dephosphorylation by CheZ). This is analogous to the case of transcription factors studied in Ref.~\cite{Burger10}: if transcription factors (or in our case CheY-P) can be degraded (or in our case dephosphorylated) only when they are not bound to their DNA targets (bound to FliM), then the concentration of non-bound transcription factors is independent of the number of DNA targets (FliM molecules). 

In Fig.~\ref{Fig5}C, $G_\textrm{CheY-P}$ features a weak maximum at about two-fold abundance of CheA. Increasing CheA abundance raises the level of phosphorylation of CheY, which yields an increase of $\psi$ (Fig.~\ref{FigS2}C), and moreover increases saturation of CheZ, which increases $G_\textrm{CheY-P}$. However, once the CheA abundance is so high that almost all CheY is phosphorylated and almost all CheZ is saturated, increasing $\alpha$ primarily increases $[\textrm{CheA-P}]$ and not $[\textrm{CheY-P}]$: hence, in this regime, the gain decreases with CheA abundance. The maximum in $G_\textrm{CheY-P}$ is smoothed out in $G_\psi$ due to FliM saturation (Fig.~\ref{Fig5}C).

Increasing CheZ abundance has the opposite effect of increasing CheA abundance, since these two enzymes have an antagonistic role in the pathway. Accordingly, $G_\textrm{CheY-P}$ features a weak maximum at about 0.5-fold abundance of CheZ (Fig.~\ref{Fig5}D), with $\psi$ decreasing when CheZ abundance increases (Fig.~\ref{FigS2}D). Increasing CheR abundance yields an increase of the active fraction $\alpha$ of CheA (Eq.~\ref{alpha}). Hence, it is effectively similar to increasing CheA abundance (Figs.~\ref{Fig5}E and~\ref{FigS2}E). Finally, increasing CheB abundance has the opposite effect, i.e. a similar effect to increasing CheZ abundance (Figs.~\ref{Fig5}F and~\ref{FigS2}F).

\section*{Discussion}

\subsubsection*{Fast signaling requirements impose strong constraints on the chemotaxis pathway}

The chemotaxis pathway is a member of the family of two-component signaling systems that enable bacteria to sense and respond to various features of their environment. This pathway is widely studied as a model signaling system. However, it faces specific constraints. Chemotaxis regulates cell swimming with response times of a fraction of a second. Longer response timescales would directly increase the lag between detection of a chemoeffector concentration and change in motion, with potentially deleterious consequences in extreme environments (e.g. in steep repellent gradients), but also in fast-changing ones. The latter case could be particularly important evolutionarily as motility peaks at the entry into stationary phase, when bacteria are competing for scarce resources~\cite{Celani10}. In contrast, the output of most other two-component systems lies in transcriptional regulation~\cite{Goulian10,Krell10,Hazelbauer10}. These systems feature overall \textit{in vivo} response times of minutes to hours~\cite{Rosenfeld02}, and their signaling involves phosphorylation reactions with \textit{in vitro} timescales of minutes~\cite{Yamamoto05}. 

We have shown that the requirements of fast signaling impose lower bounds on the dissociation constant $K_d^M$ between CheY-P and FliM, and hence on CheY-P abundance, as well as on CheZ, and CheA abundances. These lower bounds are satisfied by experimental values, giving the right order of magnitude for $K_d^M$, CheY-P, and CheZ, and indicating a low active fraction $\alpha$ of CheA in the adapted state. In practice, our pathway model gives $\alpha=0.25$ in the adapted state, within the range of previous estimates, which vary from a few percent~\cite{Sourjik02,Neumann14} to about 30\%~\cite{Sourjik02b,Shimizu10}. Note that similar constraints might exist on the abundance of CheR and CheB, since they control the dynamics of adaptation~\cite{Alon99}. However, rapidity constraints are less obvious on the adaptation timescales than on the fast response timescales studied here.

Several other features of the chemotaxis pathway reflect a pressure towards rapidity. First, the existence of a dedicated phosphatase for the response regulator CheY, which is uncommon for two-component systems, suggests the importance of fast turnover of the CheY-P pool. Second, CheA is an extremely fast histidine kinase: when incorporated in signaling complexes containing chemoreceptors and CheW, the autocatalytic rate of CheA is $k_\mathrm{cat}^A=20~/\textrm{s}$ for \textit{E. coli}~\cite{Levit02a,Francis02,Shrout03} and \textit{Salmonella typhimurium}~\cite{Levit99}, which makes it \textit{four to five orders of magnitude} faster than other kinases in two-component systems (Table~\ref{kinases}). Another possible signature of the pressure towards rapidity is that the response timescale of the chemotaxis pathway is only slightly larger than the diffusion time of CheY-P across the cytoplasm (Fig.~\ref{Fig2}B), estimated using measured diffusion coefficients~\cite{Elowitz99,Nenninger10} and a characteristic cell size of $\sim 1~\mu\textrm{m}$. Hence, the response of the chemotaxis pathway is almost as fast as it can be. Note that our model, which focuses on average concentrations, should slightly underestimate response timescales due to the neglect of diffusion. Thus, a full spatial model~\cite{Lipkow05} should yield slightly more stringent lower bounds on protein abundances.

One can wonder why the adapted CheA active fraction $\alpha$ is low ($<30\%$) while CheA is pushed towards extremely high rapidity of autophosphorylation. Ref.~\cite{Neumann14} shows that a low $\alpha$ makes the dynamics of the pathway response robust to slowly varying multiplicative noise. The pathway output is assumed to be proportional to $\alpha$, with the proportionality factor fluctuating, but more slowly than the response timescales of the pathway. In Ref.~\cite{Neumann14}, the output is chosen to be the fraction of CheZ bound to CheY-P, which is measurable by FRET. The noisy proportionality factor then involves the ratio of total CheA abundance to total CheZ abundance (see Eq.~\ref{zyp}). The robustness of the dynamics to such multiplicative noise arises from the fact that at low $\alpha$, the signal amplification at the receptor level is exponential, via the Boltzmann factor for CheA to be in its active state~\cite{Neumann14}. Since rapidity constraints imply $\alpha[\textrm{CheA}]_\textrm{tot}\gtrsim 0.25~\mu\textrm{M}$, requiring in addition an adapted $\alpha<30\%$ implies $[\textrm{CheA}]_\textrm{tot}\gtrsim 0.83~\mu\textrm{M}$, which is only $\sim 3-4$ times lower than the experimental value. Note that this would also entail a lower bound on receptor concentration of $5.0~\mu\textrm{M}$, given the stoichiometry of the array.

Since the requirement of fast signaling calls for high abundances of chemotaxis proteins, it follows that these protein levels should be higher than in homologous systems with different outputs. Many bacteria with chemotaxis pathways similar to that of \textit{E. coli}~\cite{Wuichet10}, and for which similar timescales are expected~\cite{Stocker08}, possess multiple gene clusters encoding Che proteins. Some of these paralogs regulate twitching motility based on type IV pili, while others are involved in very different cellular functions, such as development, biofilm formation, cell morphology, cell-cell interactions, and flagellar biosynthesis~\cite{Wuichet10,He14}. In the Supporting Material, we compare expression of the Che proteins involved in chemotaxis to the expression of those from paralog clusters, in five different bacteria (\textit{Pseudomonas aeruginosa}, \textit{Vibrio cholerae}, \textit{Caulobacter crescentus}, \textit{Sinorhizobium meliloti}, \textit{Rhodobacter sphaeroides}), using data from published microarray studies. We find that homologous non-chemotactic genes are significantly less expressed (at the mRNA level) than the ones actually involved in chemotaxis (Tables~\ref{PA}-\ref{RS}), with the exception of the CheY involved in twitching motility in \textit{P. aeruginosa}, which might also be subject to rapidity constraints (see Supporting Material). 

It is also interesting to compare the cellular abundances of CheA and CheY to those of the histidine kinases (HKs) and response regulators (RRs) in other two-component signaling systems. Table~\ref{twocompo} provides such a comparison for \textit{E. coli}. The protein abundance data come from several published studies and show significant variability, which may be explained by differences in media, growth phases, strains, and techniques, and the comparison should thus be taken with caution. However, it appears that CheA proteins are orders of magnitude more highly expressed than all the other HKs for which data are available (Table~\ref{twocompo}). The comparison is less striking for CheY, since it does not appear to be particularly highly expressed among RRs in the data from Ref.~\cite{Taniguchi10}, but the protein abundance measured in Ref.~\cite{Li04} is much higher, and would place CheY among the most highly expressed RRs (Table~\ref{twocompo}). While in \textit{E. coli}, CheA and CheY are expressed at comparable levels~\cite{Li04}, a number of other RRs are one or two orders of magnitude more highly expressed than their cognate HKs (Table~\ref{twocompo}). In two-component systems with bifunctional HKs that also dephosphorylate their cognate RRs, high RR abundances enable the level of phosphorylated RR to be insensitive to variations in the HK and RR abundances~\cite{Russo93,Batchelor03}. The \textit{E. coli} chemotaxis pathway is different since it possesses a dedicated phosphatase, CheZ. However, the condition for obtaining a plateau of $[\textrm{CheY-P}]$ at high abundances (Eq.~\ref{cdnFPmt}) and the corresponding adapted value of $[\textrm{CheY-P}]$ (Eq.~\ref{ypFPmt}) both depend on the ratio of CheZ to CheA abundances, which may fluctuate. Additional mechanisms provide robustness with respect to this ratio. First, CheZ is activated upon interaction with CheA-short~\cite{Wang96,OConnor04}, and most phosphatase activity takes place at the receptor arrays~\cite{Vaknin04}, which keeps phosphatase activity coupled to kinase abundance. Second, CheZ oligomerizes in the presence of CheY-P~\cite{Blat96}, and this increases its activity~\cite{Blat98}. Finally, the dependence of $\alpha$ on $[\textrm{CheB-P}]$ (Eq.~\ref{alpha}), together with the competition between CheB and CheY for CheA-P, are thought to couple kinase and phosphatase activities since CheZ and CheY abundances are strongly coupled~\cite{Neumann14}.

Hence, high chemotaxis protein abundances appear to arise from the specific rapidity constraints on the chemotaxis pathway. Supporting this view, we note that chemotaxis protein abundances similar to those in \textit{E. coli} are found in the Gram positive bacterium \textit{Bacillus subtilis}, which has even more chemoreceptors~\cite{Cannistraro11}.

\subsubsection*{Self-assembly requirements yield additional constraints}

Apart from the constraint of fast signaling, the chemotaxis pathway is also unusual among two-component systems in that it involves two types of large self-assembled multiprotein complexes: the chemoreceptor arrays, which allow for signal amplification via cooperativity~\cite{Sourjik02b,Hazelbauer07,Endres08,Hansen10}, and the rotary flagellar motors, which enable the cell to swim. The self-assembly requirements of these complexes also contribute constraints on the abundances of chemotaxis proteins.

First, inclusion in receptor arrays increases the autocatalytic rate of CheA by two orders of magnitude~\cite{Levit99}, so only CheA in arrays is functionally relevant (see Models and methods, and Table~\ref{Conc}). However, overall cellular proportions reveal a significant excess of CheA with respect to the precise 6:1:1 receptor:CheA:CheW stoichiometry of the receptor arrays~\cite{Briegel14}; for instance, overall proportions are 2.2:1:1 for strain RP437 in rich medium~\cite{Li04}. Ref.~\cite{Endres08} shows that overexpressing receptors up to $\sim 7$-fold wild-type level at native CheA level leads to a stronger response to repellent, i.e. to a stronger kinase activity, which shows that CheA is strongly in excess in these conditions too. Besides, \textit{in vitro} assembly of receptors alone leads to the formation of non-functional structures, called zippers, while adding CheA and CheW in excess to stoichiometric array proportions yields arrays~\cite{Briegel14}. Hence, in \textit{E. coli}, the correct self-assembly of the receptor arrays seems to require an excess of CheA. Note however that overall cellular proportions appear to be different in \textit{B. subtilis}, but this bacterium also expresses soluble (non-transmembrane) chemoreceptors~\cite{Cannistraro11}. 

Second, self-assembly of the flagellar motor appears to constrain the abundance of the protein FliM. In the motor, FliM forms a ring of $\sim 32$ subunits~\cite{Delalez10} which bind CheY-P to mediate switching of the direction of motor rotation. Studies~\cite{Tang95,Zhao96,Sourjik02,Delalez10} reveal that only a small fraction of FliM ($<30\%$), is part of complete motors (see Supporting Material, esp. Table~\ref{flim}). Nevertheless, underexpression and overexpression experiments indicate that FliM constitutes a limiting resource for proper motor assembly~\cite{Tang95}. Consistent with this observation, more than 25\% of FliM is found in partially assembled structures~\cite{Zhao96,Delalez10}, with only about 16\% of FliM copies free in the cytoplasm~\cite{Zhao96}. Since it is likely that FliM in partially assembled structures binds CheY-P with an affinity comparable to FliM in complete motors, these additional FliM contribute to the lower bound on the total cellular concentration of CheY-P, $C_\textrm{CheY-P}=[\textrm{CheY-P}]+[\textrm{FliM}\cdot\textrm{CheY-P}]$, yielding the second term, which accounts for 23\% of the total CheY-P lower bound. Hence, motor self-assembly requirements on FliM abundance provide a separate lower bound on CheY-P abundance only a factor of $\sim 4$ lower than our complete lower bound, which involves the actual value of $K_d^M$.

Since FliM is in excess of the requirement for complete motors, one can ask if the FliM level is constrained by signaling requirements. However, our study demonstrates that the gain, as well as the output of the pathway, are very robust to variations of the abundance of FliM (Fig.~\ref{Fig5}B). The gene encoding FliM does not belong to either of the \textit{meche} and \textit{mocha} operons that encode the Che proteins~\cite{Kollmann05,Lovdok09,Sourjik12}. Hence, FliM expression levels are likely to feature non-negligible abundance fluctuations with respect to other proteins in the pathway, making robustness to FliM abundance variations a useful feature.

\subsubsection*{Gain and cooperativity are increased by a concerted increase of protein abundances}

The abundances of chemotaxis proteins were measured in two different \textit{E. coli} strains considered wild-type for chemotaxis, in both rich and minimal growth media, in Ref.~\cite{Li04}. Strikingly, chemotaxis proteins tend to be more expressed in minimal medium than in rich medium. While this increase is modest for the reference strain RP437, where Che protein abundances increase from 1 fold to 1.1 fold, it is very strong for strain OW1, where Che protein abundances increase 9.4 times, from 0.22 fold to 2.0 fold. 

Proportions are well-conserved despite this high variability of abundances~\cite{Li04}. The Che proteins are expressed from two adjacent operons in the \textit{E. coli} genome, the \textit{meche} operon, which encodes CheR, CheB, CheY, CheZ, as well as two types of chemoreceptors, and the \textit{mocha} operon, which encodes CheA and CheW~\cite{Kollmann05,Lovdok09,Sourjik12}. Both \textit{meche} and \textit{mocha} operons are in the same regulon: they are under transcriptional control of the sigma factor $\sigma^{28}$ and of the anti-sigma factor FlgM~\cite{Kollmann05}. In addition to this transcriptional coupling, these genes also feature translational coupling~\cite{Lovdok09}. This enables the expression levels of the Che proteins to be correlated and their proportions to be stable~\cite{Kollmann05,Sourjik12}. In Ref.~\cite{Kollmann05}, where the abundances of chemotaxis proteins were varied in a concerted fashion by modulating the expression of FlgM, the chemotactic efficiency of cells (measured by a swarm assay) was found to increase sharply up to about wild-type abundance, and then to keep increasing much more gradually while progressively leveling off.

We find that the gain at the level of the response regulator CheY-P increases substantially for concerted increases of the abundances up to about reference levels, and more moderately above reference levels, reaching a plateau in the high-abundance limit. This dependence of the gain on protein abundances (Fig.~\ref{Fig4}B) is consistent with the swarm assay results of Ref.~\cite{Kollmann05}. Gain is a crucial quantity since drift velocity in a shallow chemoeffector gradient is proportional to gain~\cite{Oleksiuk11}. Moreover, an increase of the gain could further sensitize cells to small changes of attractant concentration~\cite{Sourjik02b}, which may be beneficial in poor media. This effect is strong for strain OW1, where we find that the gain in $[\textrm{CheY-P}]$ is increased by a factor $\sim 5$ in minimal medium vs. rich medium (even more if the small variations in abundance ratios~\cite{Li04} are accounted for). The small gain obtained for 0.22-fold abundance, corresponding to the expression level for strain OW1 in rich medium (Fig.~\ref{Fig4}B) arises from the small non-phosphorylated CheY reserve in this case (only $\sim 5\%$ of total CheY in adapted conditions), which entails a small response to an increase of $\alpha$. Note in addition that CheY and CheZ abundances in strain OW1 in minimal medium are smaller than our lower bounds derived from rapidity constraints, indicating slower response times. 

In addition to the increase of gain, receptor overexpression has been shown to increase cooperativity among receptors by increasing the size of receptor signaling teams~\cite{Sourjik04,Endres08,Hansen10}. This additional cooperativity can also increase sensitivity to low attractant concentrations. Together, these increases of sensitivity help explain why the proteins of the chemotaxis pathway are overexpressed in minimal medium compared to rich medium~\cite{Li04}, despite the cost of additional protein expression.

\section*{Author contributions}
Designed research: AFB and NSW; performed research: AFB; wrote the paper: AFB and NSW.

\section*{Acknowledgments}
AFB thanks Sophia Hsin-Jung Li for helpful discussions. 

This work was supported in part by National Institutes of Health Grant R01 GM082938 and National Science Foundation Grant PHY-1305525. AFB acknowledges the support of the Human Frontier Science Program.

\clearpage

\phantom{hi}
\vspace{1cm}
\centerline{\LARGE{SUPPORTING MATERIAL}}
\vspace{1cm}
\normalsize
\renewcommand{\thefigure}{S\arabic{figure}}
\setcounter{figure}{0} 
\renewcommand{\theequation}{S\arabic{equation}}
\setcounter{equation}{0} 
\renewcommand{\thetable}{S\arabic{table}}
\setcounter{table}{0} 
\renewcommand\thesubsubsection{\thesection.\arabic{subsubsection}}

\begin{spacing}{1}
\tableofcontents
\end{spacing}

\clearpage

\section{Chemotaxis pathway model: chemical reactions and parameter values}
The chemical reactions corresponding to our \textit{E. coli} chemotaxis pathway model in Eqs.~\ref{mod1}-\ref{modf} of the main text are:
\begin{align}
&\ce{\textrm{CheA} ->[\alpha k_\mathrm{cat}^A] \textrm{CheA-P}} \label{eqch1}\,,\\
&\ce{\textrm{CheA-P} + \textrm{CheY} ->[k_a^Y] \textrm{CheA} + \textrm{CheY-P}}\,,\label{eqch2}\\
&\ce{\textrm{CheA-P} + \textrm{CheB} ->[k_a^B] \textrm{CheA} + \textrm{CheB-P}}\,,\label{eqch3}\\
&\ce{\textrm{CheZ} + \textrm{CheY-P} <=>[k_a^Z][k_d^Z] \textrm{CheZ} \cdot  \textrm{CheY-P} ->[k_\mathrm{cat}^Z] \textrm{CheY} + \textrm{CheZ}}\,,\label{eqch4}\\
&\ce{\textrm{CheY-P} ->[k_h^Y] \textrm{CheY}}\,,\label{eqch5}\\
&\ce{\textrm{FliM} \cdot  \textrm{CheY-P} ->[k_h^Y] \textrm{FliM} + \textrm{CheY}}\,,\label{eqch6}\\
&\ce{\textrm{FliM} + \textrm{CheY-P} <=>[k_a^M][k_d^M] \textrm{FliM} \cdot  \textrm{CheY-P}}\,,\label{eqch7}\\
&\ce{\textrm{CheB-P} ->[k_h^B] \textrm{CheB}}\label{eqchf}\,.
\end{align}
As in the main text, phosphorylated species are denoted by ``-P'', and complexes by a dot between the two species names (e.g., $\textrm{FliM}\cdot\textrm{CheY-P}$). Eq.~\ref{eqch1} corresponds to autophosphorylation of the histidine kinase CheA, and the fraction $\alpha$ of active CheA is accounted for by an effective reduction of the autocatalytic rate from $k_\mathrm{cat}^A$ to $\alpha k_\mathrm{cat}^A$. Eq.~\ref{eqch2} and Eq.~\ref{eqch3} represent phosphotransfer from CheA-P to CheY and CheB, respectively. Eq.~\ref{eqch4} expresses dephosphorylation of the phosphorylated response regulator CheY-P by the phosphatase CheZ, and Eq.~\ref{eqch5} the (much slower) auto-dephosphorylation of CheY-P. Similarly, Eq.~\ref{eqch6} corresponds to auto-dephosphorylation of the $\textrm{FliM} \cdot  \textrm{CheY-P} $ complex, and Eq.~\ref{eqchf} to auto-dephosphorylation of CheB-P. Finally, Eq.~\ref{eqch7} represents the binding of the phosphorylated response regulator CheY-P to the FliM protein, which is a part of the flagellar motor, as well as their unbinding.

The values of the rate constants used are presented in Table~\ref{KinValues}, and the values of the effective total cellular concentrations are presented in Table~\ref{Conc}. These concentration values are used as references when abundances are varied in model calculations.
\begin{table}[htbp]
  \centering
  \caption{Values of the rate constants of the \textit{E. coli} signaling pathway used in our model. Note that a comprehensive list of experimental values is available online at `http://www.pdn.cam.ac.uk/groups/comp-cell/Data.html'. $k_a^Z$ and $k_\mathrm{cat}^Z$ were adjusted to yield consistency with FRET data (see Models and methods).}
\begin{tabular}{lll}
   \hline
   Constant&Value&Notes and references\\
   \hline
    $k_\mathrm{cat}^A$     & 20 /s &\cite{Levit02a,Francis02,Shrout03}\\
    $k_a^Y$    & 40 /s/$\mu$M &\cite{Mayover99}\\
    $k_a^B$     & 15 /s/$\mu$M &\cite{Stewart00,Pontius13}\\
    $k_a^Z$     &2.3 /s/$\mu$M & Adjusted; 5.6 /s/$\mu$M in the absence of CheA \cite{Silversmith08}.\\
    $k_d^Z$    & 0.04 /s & in the absence of CheA \cite{Silversmith08}\\
    $k_\mathrm{cat}^Z$     &12.3 /s& Adjusted; 4.9 /s  in the absence of CheA \cite{Silversmith08}.\\
    $k_h^Y$    & 0.04 /s &\cite{Silversmith01,Stewart04,Thomas08}\\
    $k_a^M$     & 5 /s/$\mu$M &Diffusion-limited~\cite{Sourjik02, Northrup92}.\\
    $k_d^M$    & 18 /s &From $K_d^M=3.5~\mu\textrm{M}$~\cite{Cluzel00,Sourjik02,Sagi03,Yuan13} and $k_a^M$.\\
    $k_h^B$    & 0.37 /s &\cite{Kentner09,Stewart93}\\
    $k_\mathrm{cat}^R$    &0.12 /s &For \textit{S. typhimurium} CheR~\cite{Simms87}.\\
    $k_\mathrm{cat}^B$    &0.29 /s &\cite{Barnakov02}\\
   \hline
\end{tabular}%
  \label{KinValues}%
\end{table}%

\begin{table}[htbp]
  \centering
  \caption{Values of the effective total cellular concentrations of the chemotaxis proteins used in our pathway model (Eqs.~\ref{mod1}-\ref{alpha}). These values derive from the total numbers of each chemotaxis protein per cell measured in Ref.~\cite{Li04} for strain RP437 in rich medium for all proteins but FliM, and from those in Refs.~\cite{Tang95,Delalez10} for FliM, also in rich medium. For CheA and FliM, we take into account additional constraints imposed by the assembly of chemoreceptor arrays and flagellar motors, respectively, as explained in Models and methods in the main text. We use the standard \textit{E. coli} cell volume of 1.4~fL~\cite{Sourjik02,Kollmann05}. }
\begin{tabular}{lll}
   \hline
   Protein&Total concentration ($\mu$M)&Notes and references\\
   \hline
CheA & 2.97 &1/6 of the chemoreceptor concentration, 17.8~$\mu$M~\cite{Li04}.\\
CheY  & 9.73 &\cite{Li04}\\
CheZ  & 3.80 &\cite{Li04}\\
CheB & 0.28 &\cite{Li04}\\
CheR  & 0.17 &\cite{Li04}\\
FliM & 1.43 &\cite{Tang95,Delalez10}. The 16\% of FliM that are free~\cite{Zhao96} are discounted. \\
   \hline
\end{tabular}%
  \label{Conc}%
\end{table}%

\clearpage

\section{Simplified pathway model}
\subsubsection{Assumptions and model}
The full pathway model, corresponding to the chemical reactions in Eqs.~\ref{eqch1}-\ref{eqchf}, is written in Eqs.~\ref{mod1}-\ref{modf} in the main text. Here, we present an analytically tractable simplified version for steady state. Our simplifying assumptions are the following:
\begin{itemize}
\item With regard to CheA-P levels, we neglect phosphotransfer to CheB with respect to phosphotransfer to CheY, i.e., we assume $k_a^Y [\textrm{CheY}]\gg k_a^B [\textrm{CheB}]$ (see Eq.~\ref{modalpha}). Indeed, the total concentration of CheY is much larger than that of CheB (Table~\ref{Conc}), and in addition $k_a^Y > k_a^B$ (Table~\ref{KinValues}). Moreover, CheY-P is dephosphorylated much faster than CheB-P, due to the existence of the dedicated phosphatase CheZ, so its turnover is much faster.
\item We treat the active fraction $\alpha$ of CheA as a parameter, without explicitly relating it to the CheB-P concentration (e.g., as in Eq.~\ref{alpha}). Thanks to this simplification, and to the previous one, CheB decouples from the rest of the system, and can thus be ignored.
\item We neglect auto-dephosphorylation of CheY-P, as it is much slower than dephosphorylation by CheZ.
\item We neglect auto-dephosphorylation of CheY-P in the complex $\textrm{FliM}\cdot\textrm{CheY-P}$, as it is much slower than dissociation of this complex ($k_h^Y\ll k_d^M$, see Table~\ref{KinValues}).
\item We neglect dissociation in the complex $\textrm{CheZ}\cdot\textrm{CheY-P}$, as it is much slower than dephosphorylation ($k_d^Z\ll k_\textrm{cat}^Z$, see Table~\ref{KinValues}).
\end{itemize}
Under these assumptions, at steady state (i.e., when all time derivatives vanish), the pathway model in Eqs.~\ref{mod1}-\ref{modf} becomes:
\begin{align}
[\textrm{CheA}]_\textrm{tot}&=[\textrm{CheA}]+[\textrm{CheA-P}]\,, \label{modS1}\\
[\textrm{CheY}]_\textrm{tot}&=[\textrm{CheY}]+[\textrm{CheY-P}]+[\textrm{FliM}\cdot\textrm{CheY-P}]+[\textrm{CheZ}\cdot \textrm{CheY-P}]\,,\\
[\textrm{CheZ}]_\textrm{tot}&=[\textrm{CheZ}]+[\textrm{CheZ}\cdot \textrm{CheY-P}]\,,\label{consZ}\\
[\textrm{FliM}]_\textrm{tot}&=[\textrm{FliM}]+[\textrm{FliM}\cdot\textrm{CheY-P}]\,,\\
\alpha k_\textrm{cat}^A [\textrm{CheA}] &= k_a^Y [\textrm{CheY}][\textrm{CheA-P}]\,, \label{modSalpha}\\
k_\textrm{cat}^Z [\textrm{CheZ}\cdot \textrm{CheY-P}] &= k_a^Y [\textrm{CheY}][\textrm{CheA-P}] \,,\label{hou}\\
k_\textrm{cat}^Z [\textrm{CheZ}\cdot \textrm{CheY-P}] &= k_a^Z [\textrm{CheZ}] [\textrm{CheY-P}]\,,\label{modSz}\\
k_a^M [\textrm{CheY-P}] [\textrm{FliM}] &= k_d^M [\textrm{FliM}\cdot\textrm{CheY-P}]\,.\label{modflimS}
\end{align}
In this system, Eqs.~\ref{modSalpha}-\ref{modSz} simply express the equality of the phosphorylation and dephosphorylation speeds of CheY at steady state.

\subsubsection{Fast phosphotransfer limit}
Given the rapidity of phosphotransfer from CheA-P to CheY at standard cellular concentrations~\cite{Mayover99}, a relevant limit is the ``fast phosphotransfer limit'', where CheA-P very rapidly undergoes phosphotransfer. In this limit,
\begin{equation}
[\textrm{CheA}]\approx[\textrm{CheA}]_\textrm{tot}\gg[\textrm{CheA-P}]\,. \label{FPeq}
\end{equation}

The simplified system Eqs.~\ref{modS1}-\ref{modflimS} can be solved analytically in the fast phosphotransfer limit, yielding successively:
\begin{align}
[\textrm{CheZ}\cdot \textrm{CheY-P}]&=\alpha \frac{k_\textrm{cat}^A}{k_\textrm{cat}^Z}[\textrm{CheA}]_\textrm{tot}\,,\label{zyp}\\
[\textrm{CheY-P}]&=\frac{\alpha \frac{k_\textrm{cat}^A}{k_a^Z}}{\frac{[\textrm{CheZ}]_\textrm{tot}}{[\textrm{CheA}]_\textrm{tot}}-\alpha\frac{k_\textrm{cat}^A}{k_\textrm{cat}^Z}}\,,\label{ypFP}\\
\psi\equiv\frac{[\textrm{FliM}\cdot\textrm{CheY-P}]}{[\textrm{FliM}]_\textrm{tot}}&=\frac{\alpha \frac{k_\textrm{cat}^A}{k_a^Z}}{\frac{k_d^M}{k_a^M}\left(\frac{[\textrm{CheZ}]_\textrm{tot}}{[\textrm{CheA}]_\textrm{tot}}-\alpha\frac{k_\textrm{cat}^A}{k_\textrm{cat}^Z}\right)+\alpha \frac{k_\textrm{cat}^A}{k_a^Z}}\label{psiFP}\,.
\end{align}
One necessary condition for the fast phosphotransfer limit to apply is that $[\textrm{CheZ}\cdot \textrm{CheY-P}]$ obtained under it (see Eq.~\ref{zyp}) should be smaller than $[\textrm{CheZ}]_\textrm{tot}$. This gives the following condition on the active fraction $\alpha$ of CheA:
\begin{equation}
\alpha<\frac{k_\textrm{cat}^Z[\textrm{CheZ}]_\textrm{tot}}{k_\textrm{cat}^A[\textrm{CheA}]_\textrm{tot}}\,.\label{cdnFP}
\end{equation}
In other words, to be in the fast phosphotransfer regime, the velocity of the autophosphorylation of CheA needs to be slower than the maximal velocity of the dephosphorylation of CheY-P by CheZ. Eqs.~\ref{ypFP}-\ref{psiFP} show that, in the fast phosphotransfer limit, both $[\textrm{CheY-P}]$ and $\psi$, which can be considered as the outputs of the pathway, depend only on the kinetic rates, on the active fraction $\alpha$ of CheA, and on the \textit{ratio} of the total concentrations of CheZ and CheA. Hence, if $\alpha$ is constant, both $[\textrm{CheY-P}]$ and $\psi$ are invariant to concerted variation of the abundances of all the proteins in the pathway, keeping the abundance ratios constant (as in Ref.~\cite{Kollmann05}). 

These results can be used to obtain the gain for the simplified pathway in the fast phosphotransfer limit. The gain in $[\textrm{CheY-P}]$, defined in Eq.~\ref{GYP}, can be obtained from Eq.~\ref{ypFP} by differentiating $[\textrm{CheY-P}]$ with respect to $\alpha$, yielding
\begin{equation}
G_\textrm{CheY-P}=\frac{\frac{[\textrm{CheZ}]_\textrm{tot}}{[\textrm{CheA}]_\textrm{tot}}}{\frac{[\textrm{CheZ}]_\textrm{tot}}{[\textrm{CheA}]_\textrm{tot}}-\alpha\frac{k_\textrm{cat}^A}{k_\textrm{cat}^Z}}\,.\label{GS1}
\end{equation}
Using Eqs.~\ref{consZ} and~\ref{zyp}, we can express this gain as
\begin{equation}
G_\textrm{CheY-P}=\frac{[\textrm{CheZ}]_\textrm{tot}}{[\textrm{CheZ}]}\,.\label{GS1b}
\end{equation}
This expression demonstrates that the gain in $[\textrm{CheY-P}]$ arises from saturation of the phosphatase CheZ. Similarly, the gain in $\psi$, defined in Eq.~\ref{Gpsi}, can be obtained from Eq.~\ref{psiFP} by differentiating $\psi$ with respect to $\alpha$, yielding
\begin{equation}
G_\psi=\frac{\frac{[\textrm{CheZ}]_\textrm{tot}}{[\textrm{CheA}]_\textrm{tot}}}{\frac{[\textrm{CheZ}]_\textrm{tot}}{[\textrm{CheA}]_\textrm{tot}}+\alpha\left(\frac{k_\textrm{cat}^A}{k_a^Z}\frac{k_a^M}{k_d^M}-\frac{k_\textrm{cat}^A}{k_\textrm{cat}^Z}\right)}=\frac{[\textrm{CheZ}]_\textrm{tot}}{[\textrm{CheA}]_\textrm{tot}}\frac{\psi}{\alpha}\frac{k_a^Z}{k_\textrm{cat}^A}\frac{k_d^M}{k_a^M}\,. \label{GS2}
\end{equation}

\subsubsection{Comparison with results from the full pathway}
The validity of the fast phosphotransfer limit in Eq.~\ref{FPeq}, and of the assumptions in our simplified pathway model (see above), can be tested against the results from the full pathway. We find that the results agree well at reference expression levels and higher of the chemotaxis signaling proteins, for the parameter values used here (kinetic rates in Table~\ref{KinValues} and abundance ratios equal to those in Table~\ref{Conc}). 

\begin{figure}[h t b]
\centering
\includegraphics[width=0.45\textwidth]{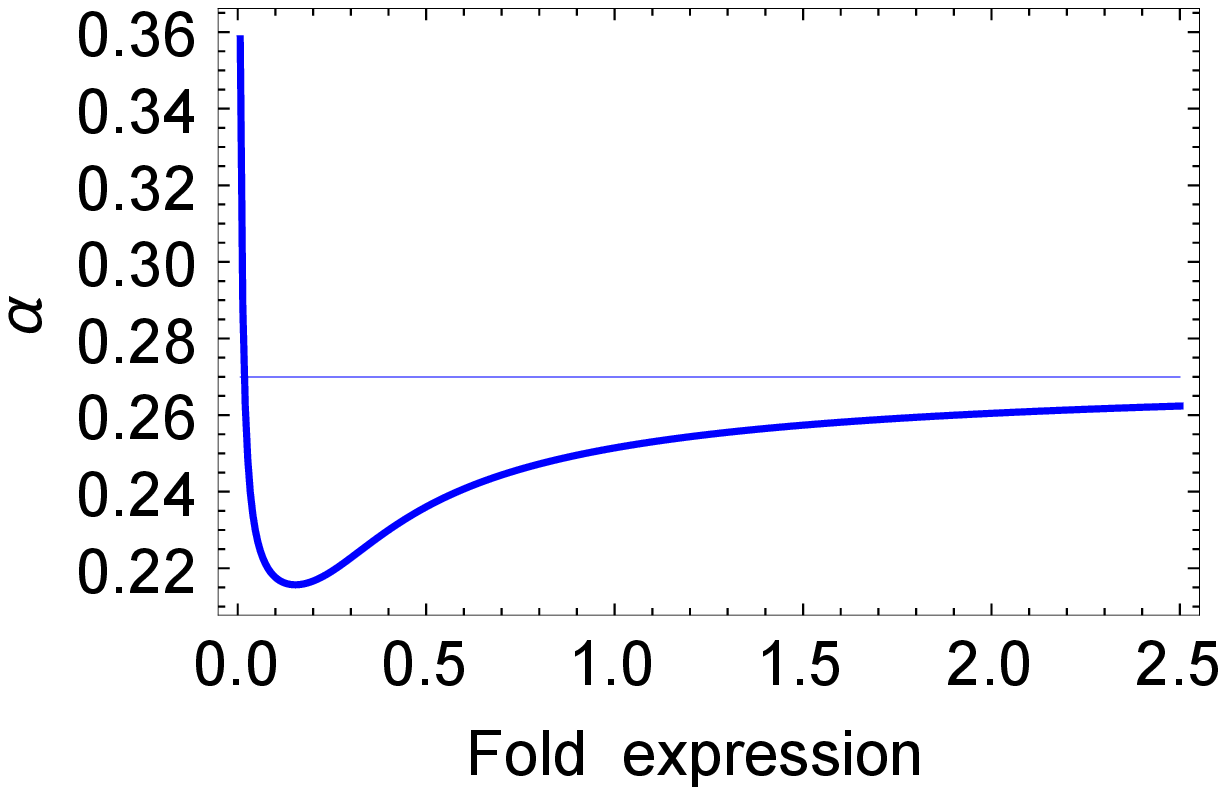}
\caption{\label{FigS1}Adapted active fraction $\alpha$ of CheA versus fold expression of all chemotaxis signaling proteins, obtained from the pathway model in the adapted state (Eqs.~\ref{mod1}-\ref{alpha}). Thick curve: result from the full pathway model. Thin line: asymptotic high-abundance result from the simplified pathway model in the fast phosphotransfer regime. One-fold expression corresponds to the abundances measured in Ref.~\cite{Li04} for strain RP437 in rich medium. In the same rich medium, the fold expression for strain OW1 is about 0.22, while in minimal medium, the fold expression is about 1.1 for strain RP437 and 2.0 for strain OW1~\cite{Li04} (values averaged over all the chemotaxis signaling proteins). See also Fig.~\ref{Fig4} in the main text. }
\end{figure}

In our full pathway model, $\alpha$ is coupled to the rest of the pathway through $[\textrm{CheB-P}]$ (see Eq.~\ref{alpha}). Solving the full pathway yields $\alpha$ as a function of the expression level of all chemotaxis signaling proteins (Fig.~\ref{FigS1}). At high expression levels, $\alpha$ reaches a plateau. The asymptotic value of $\alpha$ at high expression levels can be calculated within the fast-phosphotransfer limit of our simplified pathway. For this, we express $[\textrm{CheB-P}]$ as a function of $\alpha$,  using Eq.~\ref{consb} and Eq.~\ref{modf} at steady state, and use the solutions of the simplified pathway in the fast phosphotransfer limit for $[\textrm{CheA-P}]$. We then use Eq.~\ref{alpha} together with this expression for $[\textrm{CheB-P}]$ in order to solve for $\alpha$. In the limit of high abundances (keeping abundance ratios constant), this amounts to solving a second-degree equation, which yields the asymptotic value of $\alpha$. With the parameter values used here (kinetic rates in Table~\ref{KinValues} and abundance ratios equal to those in Table~\ref{Conc}), we obtain $\alpha=0.27$, close to the value $\alpha=0.26$ obtained for 2.5-fold overexpression from the full pathway model (Fig.~\ref{FigS1}). 

Hence, the results from our simplified pathway model in the fast phosphotransfer limit are relevant at high abundances, and account for the observed plateaus of $[\textrm{CheY-P}]$ and $\psi$ in Fig.~\ref{Fig4}A, and of the gain in Fig.~\ref{Fig4}B. Using the high-abundance asymptotic value $\alpha=0.27$, Eq.~\ref{ypFP} yields $[\textrm{CheY-P}]=2.8~\mu\textrm{M}$, and Eq.~\ref{psiFP} yields $\psi=0.44$. These values are close to those obtained for high overexpression (specifically, $[\textrm{CheY-P}]=2.6~\mu\textrm{M}$, and $\psi=0.43$ for 2.5-fold overexpression, see Fig.~\ref{Fig4}A). Similarly, Eq.~\ref{GS1} yields $G_\textrm{CheY-P}=1.5$ here, and Eq.~\ref{GS2} yields $G_\psi=0.84$, extremely close to the values obtained for high overexpression (Fig.~\ref{Fig4}B). These asymptotic high-abundance values from our simplified pathway model in the fast phosphotransfer limit are plotted as thin lines in Fig.~\ref{Fig4}A-B. In addition, the CheZ concentration from the full pathway solution is used to plot the ratio in Eq.~\ref{GS1b}, which is the dotted line in Fig.~\ref{Fig4}B. It gives a good approximation to the actual gain in $[\textrm{CheY-P}]$ for sufficiently high abundances, of order one-fold and above.

\subsubsection{Low-abundance limit}
While the fast phosphotransfer regime is the relevant one for standard cellular abundances and higher, the pathway's behavior is very different in the limit of low abundances. Indeed, since autophosphorylation of CheA is autonomous, while phosphotransfer to CheY involves a two-molecule encounter between CheA-P and CheY, the fraction of phosphorylated CheA becomes high in the limit of low overall abundances. More precisely, if $[\textrm{CheY}]_\textrm{tot}\ll \alpha k_\textrm{cat}^A/k_a^Y$, then Eq.~\ref{modSalpha} ensures that 
\begin{equation}
[\textrm{CheA}]\ll[\textrm{CheA-P}]\approx[\textrm{CheA}]_\textrm{tot}\,. 
\end{equation}
Provided that $[\textrm{CheY}]_\textrm{tot}\ll k_\textrm{cat}^Z/k_a^Z$ and that $[\textrm{CheY}]_\textrm{tot}\ll K_d^M\equiv k_d^M/k_a^M$, small proportions of CheZ and of FliM are bound to CheY-P. We obtain, using Eqs.~\ref{modSz} and~\ref{modflimS},
\begin{align}
[\textrm{CheZ}\cdot \textrm{CheY-P}] &\approx\frac{k_a^Z}{k_\textrm{cat}^Z } [\textrm{CheZ}]_\textrm{tot} [\textrm{CheY-P}]\,,\label{modSzb}\\
[\textrm{FliM}\cdot\textrm{CheY-P}]&\approx \frac{k_a^M}{k_d^M } [\textrm{FliM}]_\textrm{tot} [\textrm{CheY-P}] \,.\label{modflimb}
\end{align}
In this regime, if in addition $[\textrm{CheZ}]_\textrm{tot}\ll k_\textrm{cat}^Z/k_a^Z$, whic implies that a small fraction of CheY-P is bound to CheZ (see Eq.~\ref{modSzb}), Eqs.~\ref{hou} and~\ref{modSz} yield
\begin{equation}
[\textrm{CheY-P}]\approx\frac{[\textrm{CheY}]_\textrm{tot}}{1+\frac{k_a^M}{k_d^M}[\textrm{FliM}]_\textrm{tot}+\frac{k_a^Z}{k_a^Y}\frac{[\textrm{CheZ}]_\textrm{tot}}{[\textrm{CheA}]_\textrm{tot}}}\,. \label{sollow}
\end{equation}
Given that we vary the overall abundances of CheA, CheY, CheZ, CheB, and CheR, while keeping their proportions and the FliM abundance fixed (see main text), Eq.~\ref{sollow} shows that in the low-abundance limit, $[\textrm{CheY-P}]$ grows in proportion to the overall abundances of the Che proteins (indeed it shows that $[\textrm{CheY-P}]\propto[\textrm{CheY}]_\textrm{tot}$).

Eqs.~\ref{modflimb} and~\ref{sollow} yield an expression for $\psi$ in this limit:
\begin{equation}
\psi\approx\frac{[\textrm{CheY-P}]}{K_d^M}\approx\frac{[\textrm{CheY}]_\textrm{tot}}{K_d^M+[\textrm{FliM}]_\textrm{tot}+K_d^M\frac{k_a^Z}{k_a^Y}\frac{[\textrm{CheZ}]_\textrm{tot}}{[\textrm{CheA}]_\textrm{tot}}}\,. \label{sollowb}
\end{equation}
These asymptotic low-abundance expressions from our simplified pathway model are plotted as thin dotted lines in Fig.~\ref{Fig4}A. 

Note that here, we have just studied the low-abundance limit of our simplified pathway model. In practice, additional effects might come into play, for instance the formation of the chemoreceptor array and of the flagellar motor would likely be impaired at too low abundances.

\clearpage

\section{Effect of a variation of the level of each protein of the pathway}

\begin{figure}[h t b]
\centering
\includegraphics[width=0.6\textwidth]{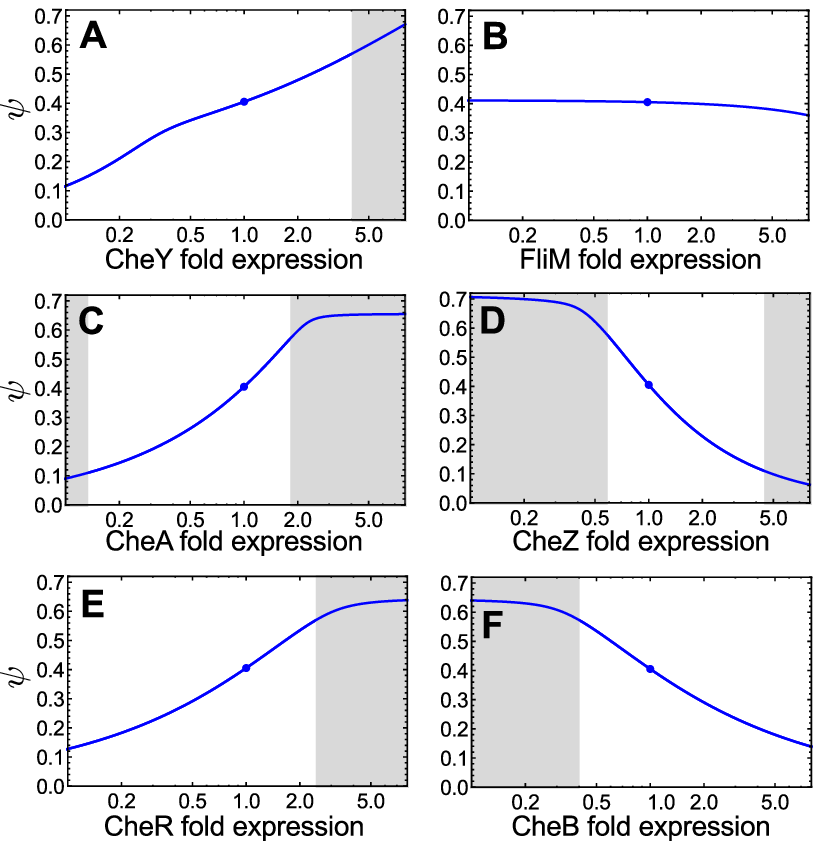}
\caption{\label{FigS2}Effect of fold-change of expression of each chemotaxis signaling protein separately, obtained from the pathway model in the adapted state (Eqs.~\ref{mod1}-\ref{alpha}). One-fold expression corresponds to the abundances in Table~\ref{Conc}, i.e. to those measured in Ref.~\cite{Li04} for strain RP437 in rich medium, as in Fig.~\ref{Fig4}. \textbf{A}-\textbf{F.}  Blue curves: adapted fraction $\psi$ of FliM proteins bound to CheY-P versus fold expression of each protein, keeping all others at their one-fold level. Blue dots: one-fold expression case.  In the shaded zones, $\psi$ is either smaller than 0.11 or larger than 0.57, in which case the flagellar motor should rotate only counterclockwise or only clockwise, respectively, in the adapted state~\cite{Sourjik02}.}
\end{figure}

\begin{figure}[h t b]
\centering
\includegraphics[width=0.6\textwidth]{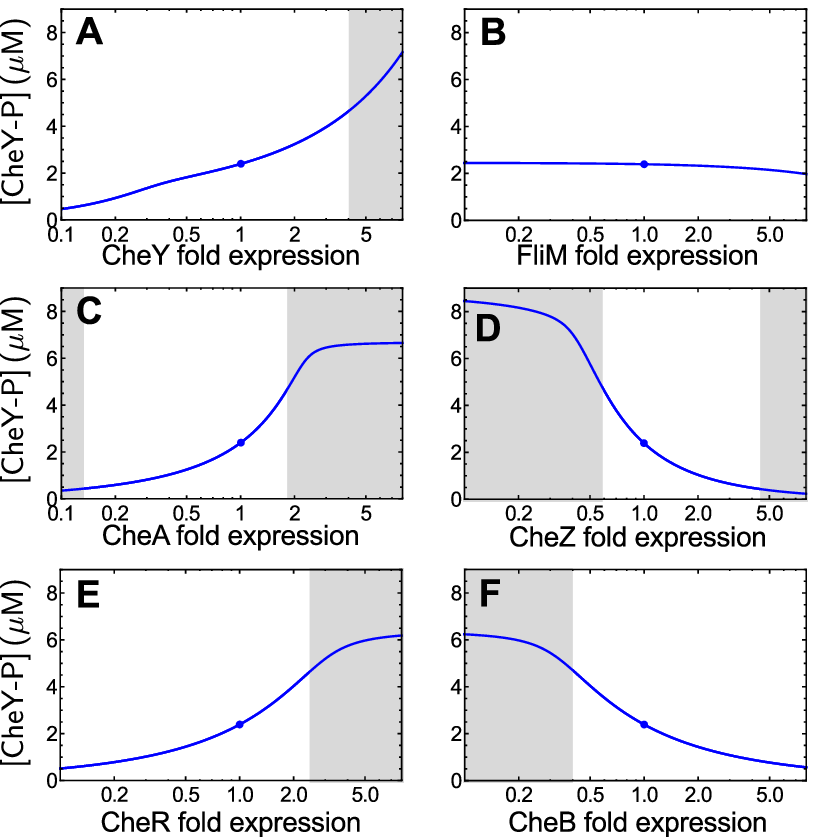}
\caption{\label{FigS2_yp}Adapted free CheY-P concentration, $[\textrm{CheY-P}]$, versus fold expression of each protein, keeping all others at their one-fold level. Same conventions as in Fig.~\ref{FigS2}. Here too, in the shaded zones, $\psi$ is either smaller than 0.11 or larger than 0.57 (see Fig.~\ref{FigS2}).}
\end{figure}

The effect of varying the level of each protein of the chemotaxis pathway is discussed in the main text, and Fig.~\ref{Fig5} shows how the gain is affected by these individual variations of protein levels. Here, we present results regarding the direct outputs of the pathway, namely the adapted fraction $\psi$ of FliM proteins bound to CheY-P (Fig.~\ref{FigS2}) and the adapted free CheY-P concentration, $[\textrm{CheY-P}]$ (Fig.~\ref{FigS2_yp}). 

These results, especially Fig.~\ref{FigS2_yp}, enable a direct comparison with Ref.~\cite{Levin98}, where $[\textrm{CheY-P}]$ was considered, but not the gain of the pathway. 

\clearpage
\section{Comparison of the autocatalytic rates of different histidine kinases}

Table~\ref{kinases} lists experimental values from the literature for the autocatalytic rates of different histidine kinases involved in bacterial two-component signaling systems. CheA is much faster than all of these, with an autocatalytic rate $k_\textrm{cat}^A=2.6\times 10^{-2}~/\textrm{s}$ in \textit{E. coli}~\cite{Tawa94} and $k_\textrm{cat}^A=0.24~/\textrm{s}$ in \textit{S. typhimurium}~\cite{Levit99} when isolated, and $k_\mathrm{cat}^A=23~/\textrm{s}$ in \textit{S. typhimurium}~\cite{Levit99} when in complex with chemoreceptors and CheW.

\begin{table}[htbp]
  \centering
  \caption{Autocatalytic rate $k_\textrm{cat}$ for various histidine kinases (from different bacteria). }
    \begin{tabular}{lllll}
   \hline
  Organism & Function & Histidine kinase   & $k_\textrm{cat}$ (/s)    & Ref. \\
    \hline
 \textit{Enterococcus faecium}  & Antibiotic resistance& VanS  & $2.83\times 10^{-3}$ & \cite{Wright93} \\
  \textit{Bacillus subtilis}  & Sporulation & KinA& $1.90\times 10^{-3}$ &  \cite{Grimshaw98}\\
 \textit{Bacillus subtilis} & Cold shock response & DesK  & $2.80\times 10^{-3}$ & \cite{Trajtenberg10}\\
  \textit{Thermotoga maritima } &  &   HpkA  & $4.23\times 10^{-4}$ &  \cite{Foster04}\\
  \textit{Streptococcus pneumoniae} & Virulence, etc. &  WalKSpn (C)-His & $1.40\times 10^{-3}$ &  \cite{Gutu10}\\
  \textit{Streptococcus pneumoniae} & Virulence, etc. &  WalKSpn (N)-Sumo & $3.60\times 10^{-3}$ &  \cite{Gutu10}\\
  \textit{Escherichia coli} & Response to nitrite &   NarX & $5.00\times 10^{-5}$ &  \cite{Noriega08}\\
  \textit{Escherichia coli} & Response to nitrate & NarQ & $2.20\times 10^{-4}$ &  \cite{Noriega08}\\
  \textit{Synechocystis} & Light signaling system &  Cph1 holo & $2.00\times 10^{-4}$ &  \cite{Psakis11}\\
  \textit{Synechocystis} & Light signaling system &  Cph1 apo & $3.00\times 10^{-4}$ &  \cite{Psakis11}\\
   \textit{Myxococcus xanthus} & Aggregation; sporulation &  RodK & $1.67\times 10^{-4}$ &  \cite{Rasmussen06}\\
    \hline
    \end{tabular}%
  \label{kinases}%
\end{table}

\newpage

\section{Expression levels of various paralogs of the chemotaxis gene clusters}
\setcounter{subsubsection}{0}

Here, we compare the expression levels of \textit{che} genes actually involved in chemotaxis to those of non-chemotactic \textit{che} genes in bacteria that have multiple \textit{che} gene clusters in their genome. For this comparison, we use data from published microarray studies.

\subsubsection{\textit{Pseudomonas aeruginosa}}

The genome of \textit{P. aeruginosa} includes four main clusters of \textit{che} genes~\cite{Wuichet10}. Among these, two are involved in chemotaxis, with one, PA1456-1464 (+ PA3348-3349 containing \textit{cheR}), being essential for chemotaxis, and the second one (PA0180-0173) being non-essential. Ref.~\cite{Hong04} showed that strains deleted for the latter cluster exhibit positive chemotactic response (to peptone and phosphate). In Ref.~\cite{Ferrandez02}, this cluster was found to be required for an optimal chemotactic response (in addition to the main cluster), but the authors state that they cannot exclude that its major output is some other function.  Among the two remaining clusters, one (PA0408-0415) is involved in twitching motility based on type IV pili~\cite{Whitchurch04}, and the last one (PA3708-3702) regulates biofilm formation through modulation of c-diGMP levels~\cite{Hickman05}. 

In Ref.~\cite{Chang05}, a microarray analysis of \textit{P. aeruginosa} was conducted to study its response to hydrogen peroxide. The full microarray data are available both in the absence (control) and in the presence of hydrogen peroxide. Table~\ref{PA} corresponds to the microarray results (i.e., the abundance of mRNA) in the control data, for \textit{cheY}, \textit{cheA}, \textit{cheW}, \textit{cheR}, and \textit{cheB} from all chemotaxis clusters. It gives the ratio of the expression levels of the genes from the main chemotaxis cluster to those of the corresponding genes from each other cluster.

\begin{table}[h t b p]
  \centering
  \caption{Microarray expression data for \textit{P. aeruginosa}, from controls in Ref.~\cite{Chang05}. The numbers given are ratios of the expression levels of the genes from each of the three clusters of \textit{che} genes that are non-essential to chemotaxis, to those of the corresponding genes from the essential chemotaxis cluster, PA1456-1464 (+ PA3348-3349 containing \textit{cheR}). The indication ``(2)'' means that two genes of this type are present in the cluster considered (two \textit{cheW} genes exist in the main chemotaxis cluster, as well as in the twitching-associated one and in the biofilm-associated one, and two \textit{cheY} genes exist in the twitching-associated cluster). In these cases, the results obtained for each of the gene copies within the cluster considered were summed.\protect\\
*Note that the response regulator encoded by the biofilm-associated cluster, WspR, is classified as non-CheY~\cite{Wuichet10}. }
\begin{tabular}{llll}
\hline
&Chemotaxis II&Twitching&Biofilm formation\\
& PA0180-0173&PA0408-0415&PA3708-3702\\
\hline
\textit{cheY}&0.22&\textbf{5.34 (2)}&0.22*\\
\textit{cheA}&0.09&0.93&0.23\\
\textit{cheW} (2)&0.02&0.58 (2)&0.24 (2)\\
\textit{cheR}&0.15&0.52&0.21\\
\textit{cheB}&0.02&0.30&0.31\\
\hline
\end{tabular}
  \label{PA}%
\end{table}%

Table~\ref{PA} shows that genes from the main chemotaxis cluster are significantly more highly expressed than those of the second cluster involved in chemotaxis. They are also more expressed than genes from the biofilm-associated one. However, the \textit{cheA} and \textit{cheW} genes from the twitching-associated cluster have similar levels of expression as those of the main chemotaxis cluster, and the two copies of \textit{cheY} in the twitching-associated cluster are together five times more expressed than the \textit{cheY} involved in chemotaxis. This twitching motility-associated \textit{che} gene cluster is known to modulate the activity of an adenylate cyclase involved in virulence, and to have additional roles in transcriptional regulation, but this pathway is complex and not fully elucidated yet~\cite{He14}. Chemosensing for directed twitching motility (on surfaces) has to be fast as in swimming, and the twitching-motility system is known to mediate chemotaxis towards phospholipids~\cite{Miller08}. Hence, the abundance of these motility-associated proteins might partly arise from rapidity constraints, as in the case of swimming chemotaxis. In addition, Ref.~\cite{Whitchurch04} suggests a possible role of (one of) the CheYs of this system as a phosphate sink (as in the chemotaxis systems of \textit{Rhodobacter sphaeroides}, which include no CheZ): this might explain the very high expression of the two \textit{cheY} genes in the \textit{P. aeruginosa} twitching motility-associated cluster. 

Notably, the main chemotaxis cluster is the only one that includes CheZ, whose role is to dephosphorylate CheY rapidly.

\subsubsection{\textit{Vibrio cholerae}}

The genome of \textit{V. cholerae} includes three main clusters of \textit{che} genes~\cite{Wuichet10}. Among these, only one, VC2059-2065 (+ VC2201-2202 containing \textit{cheR}) is involved in chemotaxis. The functions of the other two clusters (VC1394-1406 and VCA1088-1096) are not known yet. In fact, neither deletion nor overexpression of the genes in these clusters has been found to produce any detectable phenotype~\cite{Nishiyama12}.

In Ref.~\cite{Xu03}, the transcriptome of \textit{V. cholerae} was studied both for bacteria grown \textit{in vitro} and during intraintestinal growth (in the latter case, the bacteria were harvested from rabbit ileal loops). The full microarray data are available in both of these conditions. Table~\ref{VC}(a) and (b) gives the ratio of the expression levels of the genes from the actual chemotaxis cluster to those of the corresponding genes from each other cluster, calculated from the microarray data of Ref.~\cite{Xu03}. 

\begin{table}[htbp]
  \centering
  \caption{Similar data as in Table~\ref{PA}, for \textit{V. cholerae}. (a) and (b): Microarray expression data from controls in Ref.~\cite{Xu03} -- (a): harvested from rabbit ileal loops; (b): grown \textit{in vitro}. }

\begin{tabular}{lll}
\hline
\textbf{(a)}&VC1394-1406&VCA1088-1096\\
\hline
\textit{cheY}&0.45 (2)&0.29\\
\textit{cheA}&0.70&0.25\\
\textit{cheW}&0.30&0.53 (2)\\
\textit{cheR}&0.63&0.62\\
\textit{cheB}&0.14&0.20\\
\hline
\end{tabular}

\begin{tabular}{lll}
\hline
\textbf{(b)}&VC1394-1406&VCA1088-1096\\
\hline
\textit{cheY}&0.22 (2)&0.11\\
\textit{cheA}&0.89&0.10\\
\textit{cheW}&0.16&0.42 (2)\\
\textit{cheR}&0.33&0.44\\
\textit{cheB}&0.14&0.24\\
\hline
\end{tabular}

  \label{VC}%
\end{table}%

In spite of some variability between conditions, the data in Table~\ref{VC} consistently show that the actual chemotaxis cluster is more expressed than the other two \textit{che} gene clusters. Here too, the actual chemotaxis cluster is the only one that includes CheZ, whose role is to dephosphorylate CheY rapidly.

\subsubsection{\textit{Caulobacter crescentus}}

\textit{C. crescentus} has two different \textit{che} gene clusters, and only one of them is known to be involved in chemotaxis. Here we compare the expression levels of the second \textit{che} cluster (CCNA00625-00634) to those of the genes involved in chemotaxis (CCNA00439-00450). Ref.~\cite{Fang13} investigated gene expression in different phases of the cell cycle of \textit{C. crescentus}, and reported a full microarray study of \textit{C. crescentus} in these different phases.

\begin{table}[htbp]
  \centering
  \caption{Similar data as in Table~\ref{PA}, for \textit{C. crescentus}. Microarray expression data from Ref.~\cite{Fang13}. Phases of the cell cycle: swarmer (SW), stalk (ST), early predivisional (EPD), predivisional
(PD), and late predivisional (LPD).}
    \begin{tabular}{llllll}
    \hline
    &CCNA00625-00634&&&&\\
       &SW & ST & EPD & PD & LPD \\
\hline
\textit{cheY} (3)&0.24 (3)&1.17 (3)&0.10 (3)&0.06 (3)&0.14 (3)\\
\textit{cheA}&Below detection level&Below detection level&0.03&0.03&0.04\\
\textit{cheW}&0.17&0.11&0.06&0.06&0.06\\
\textit{cheR}&0.27&1.00&0.13&0.06&0.14\\
\textit{cheB}&0.02&Below detection level&0.01&0.02&0.02\\
\hline
\end{tabular}

  \label{CC}%
\end{table}%

\subsubsection{\textit{Sinorhizobium meliloti}}

\textit{S. meliloti} has two different \textit{che} gene clusters, and only one of them is known to be involved in chemotaxis. There are also two isolated \textit{cheW} genes (one being close to an MCP-coding gene). Here we compare the expression levels of the second chemotaxis cluster (SMa1550-1561) to those of the genes of the main cluster involved in chemotaxis (SMc03004-03012). We do not consider the isolated \textit{cheW} gene because it is not known whether it is involved in chemotaxis. Ref.~\cite{Tiricz13} investigated the influence of antimicrobial nodule-specific cysteine-rich peptides, and reported full microarray study of \textit{S. meliloti}.

\begin{table}[htbp]
  \centering
  \caption{Similar data as in Table~\ref{PA}, for \textit{S. meliloti}. Microarray expression data from controls in Ref.~\cite{Tiricz13} (10 or 30 min incubation with shaking in modified LSM medium, after the addition of sterile water -- instead of antimicrobial peptide).\protect\\
  *Note that the response regulator encoded by the second cluster, CheO, is classified as non-CheY~\cite{Wuichet10}.}

      \begin{tabular}{lll}
    \hline
    & SMa1550-1561 (10 min)&SMa1550-1561 (30 min)\\
    \hline
\textit{cheY} (2)&0.52$^*$&0.78$^*$\\
\textit{cheA}&0.33&0.31\\
\textit{cheW}&0.30&0.27\\
\textit{cheR}&0.54&0.57\\
\textit{cheB}&0.30&0.27\\
\hline
\end{tabular}

  \label{SM}%
\end{table}%

\subsubsection{\textit{Rhodobacter sphaeroides}}

\textit{R. sphaeroides} has three different \textit{che} gene clusters, and two of them are known to be involved in chemotaxis. There is in addition one isolated \textit{cheY}, which is also essential to chemotaxis. Here we compare the expression levels of the third \textit{che} cluster to those of the genes involved in chemotaxis. Ref.~\cite{Peuser12} investigated the role of a protein in the iron metabolism of \textit{R. sphaeroides}, and reported full microarray studies of \textit{R. sphaeroides} in different conditions. Here we report the results obtained for the wild-type strain in the presence and in the absence of iron. Similarly, Ref.~\cite{Metz12} used microarrays to investigate the role of a light, oxygen, voltage domain protein in blue light-dependent and singlet oxygen-dependent gene regulation in \textit{R. sphaeroides} and reported full microarray data.

Note that these data are to be interpreted with caution, first because of the complication of \textit{R. sphaeroides} having two essential chemotaxis systems, and second because the expression level of many of the genes considered here were found to be below the threshold of significance defined in the original publications. Note also that the non-chemotactic \textit{che} cluster encodes a putative CheX, and that CheX plays the part of a CheY phosphatase in \textit{Borrelia burgdorferi}~\cite{Motaleb05}.

\begin{table}[htbp]
  \centering
  \caption{Similar data as in Table~\ref{PA}, for \textit{R. sphaeroides}. (a): Microarray expression data from controls in Ref.~\cite{Peuser12}. (b): Microarray expression data from controls in Ref.~\cite{Metz12}.}
    \begin{tabular}{lll}
    \hline
      \textbf{(a)}    & RSP2433-2443 (WT+Fe) & RSP2433-2443 (WT$-$Fe) \\
    \hline
    \textit{cheY} (3)    &1.40 (3) & 0.53 (3) \\
    \textit{cheA} (3)    & 0.04  & 0.01 \\
    \textit{cheW} (3)   & Below detection level  & 0.04 \\
    \textit{cheR} (2)    & 0.09  & 0.25 \\

   \hline
    \end{tabular}%

      \begin{tabular}{ll}
    \hline
      \textbf{(b)}    & RSP2433-2443 (WT)\\
    \hline
    \textit{cheY} (3)    & 0.92 (3)\\
    \textit{cheA} (3)    & 0.34\\
    \textit{cheW} (3)   & 0.13 \\
    \textit{cheR} (2)    & 0.30 \\
   \hline
    \end{tabular}%

  \label{RS}%
\end{table}%

\newpage

\section{Expression levels of genes coding for two-component systems in \textit{E. coli} }

The genome of \textit{E. coli} comprises 29 sensor histidine kinases and 32 response regulators involved in two-component signaling systems~\cite{Oshima02,Krell10}, which are involved in the cell's sensing and response to its environment. It is interesting to compare the expression levels of the chemotaxis genes to those of other two-component signaling systems. In Table~\ref{twocompo}, we present the number of protein copies per cell for the histidine kinases and response regulators of various two-component systems in \textit{E. coli}. The variability across estimates from different studies may be explained by the differences in media, growth phases, and strains of \textit{E. coli}, which are known to yield significant variations of abundances~\cite{Li04,Cai02,Ishihama08}, and also perhaps by the different techniques used:
\begin{itemize}
\item In Ref.~\cite{Taniguchi10}, the proteome of \textit{E. coli} was quantified at the single-cell level using single-molecule fluorescence, thanks to a yellow-fluorescent-protein fusion library. \textit{E. coli} BW25993 cells were grown in LB media and then inoculated into M9 media supplemented with glucose, amino acids, and vitamins. The optical density was 0.1-0.5.
\item Ref.~\cite{Ishihama08} used mass spectrometry (more precisely, emPAI) to quantify the abundance of proteins in \textit{E. coli}, focusing mostly on cytoplasmic proteins. \textit{E. coli} MC4100 cells were grown in rich or minimum medium to exponential phase (optical density $\sim$0.4), but the datasets obtained with the two different media were combined in the final analysis.
\item Ref.~\cite{Masuda09} used the emPAI technique as Ref.~\cite{Ishihama08}, but this work also quantified the abundances of membrane proteins, first extracting them with the aid of a removable phase transfer surfactant (PTS). In this reference, \textit{E. coli} BW25113 cells were grown in LB medium and harvested at stationary phase.
\item Ref.~\cite{Li04} used quantitative immunoblotting, focusing only on chemotaxis proteins. Several conditions were studied. Here we report the values obtained for strain RP437 in Tryptone broth (rich medium) with cells grown to an optical density of 0.5.
\item Ref. \cite{Cai02} focused on the levels of EnvZ and OmpR proteins using quantitative Western blot analysis. The \textit{E. coli} strain used was MC4100. The values reported here were obtained during exponential growth in L-broth medium.
\end{itemize}

\begin{table}[htbp]
  \centering
  \caption{Cellular abundances of the proteins from various two-component systems in \textit{E. coli}. Data from several proteomic studies (see also `http://ecoliwiki.net/colipedia'). All values are in copy numbers per cell.}
  \footnotesize
    \begin{tabular}{l|llll|llll}
   \hline
Two-component system &\multicolumn{4}{l|}{Histidine kinase} & \multicolumn{4}{l}{Response regulator}\\
(Histidine kinase / response regulator) &\cite{Taniguchi10} &\cite{Ishihama08} & \cite{Masuda09}&Other & \cite{Taniguchi10} & \cite{Ishihama08}& \cite{Masuda09} & Other\\
    \hline
CheA/ CheY: Chemotaxis & & & &6,700~\cite{Li04}&12&  & & 8,200~\cite{Li04}\\
EnvZ / OmpR: Osmolarity sensing& & &1.3&100~\cite{Cai02}&81 & 613 & 238& 3,500~\cite{Cai02}\\ 
NarX / NarL: Response to nitrite& & &1.3& & &229 &522&\\ 
PhoR / PhoB: Phosphate regulation&11& & & & & & & \\ 
EvgS / EvgA: Drug resistance&5&82& & & &198&26& \\ 
CusS / CusR: Copper response&2& &1.4& & & & & \\ 
YedV / YedW&3& &0.8& &1.5& & & \\ 
KdpD / KdpE: Potassium transport&9& &6/0.7& &6& & & \\ 
BaeS / BaeR& & & & &12&167& & \\ 
HydH / HydG&1& & & & & & &  \\ 
PhoQ / PhoP: Response to magnesium&7& &1.2& & &786&113& \\ 
BasS / BasR: Polymyxin resistance&0.7& &1.3& &65& & & \\ 
CpxA / CpxR: Response to cell envelope stress& & &1.8& &33&664&316& \\ 
TorS / TorR&3& & & & & & & \\ 
DcuS / DcuR& & &1.1 & &0.6& & &\\ 
RcsC / RcsB: Capsular synthesis&9& &1.1& &369&1,490&597& \\ 
CitA / CitB& & & & &1& & &\\ 
ArcB / ArcA: Respiratory control&56&100&32/1.5& & &2,660&550& \\ 
BarA / UvrY: Hydrogen peroxide sensitivity&2& &1.3& &29& & 18&\\
    \hline
    \end{tabular}%
    \normalsize
  \label{twocompo}%
\end{table}

\newpage

\section{Cellular localization of the protein FliM}

The protein FliM is a constituent of the cytoplasmic ring of the rotor that mediates rotation-direction switching in response to binding of CheY-P. Several studies reveal that only a small fraction of FliM is part of complete motors (Table~\ref{flim}): 
\begin{itemize}
\item In Ref.~\cite{Tang95}, the relative abundances of FliM in the membrane and cytoplasmic fractions were estimated by lysing the cells and separating the membranes from the cytoplasm by centrifugation. It was found that $\sim$1,100 copies of FliM out of 1,400, i.e., $\sim$78\% of FliM copies, were in the cytoplasmic fraction, and thus not in complete motors.
\item In Ref.~\cite{Delalez10}, the abundance and localization of FliM were studied by fluorescence microscopy. There were 24 $\pm$ 6 spots, corresponding to assemblies of multiple FliM, per cell. The distribution of the number of molecules per spot showed two peaks, one at 32 molecules (in agreement with previously measured numbers of FliM per flagellar motor) accounting for about 40\% of the spots, and one at 18 molecules, which may correspond to partly assembled cytoplasmic rings, accounting for about 45\% of the spots~\cite{Delalez10}. The remaining $~15\%$ of the spots fell outside these peaks, corresponding to structures not independently resolved. Hence, $40-55\%$ of the FliM in spots were part of full cytoplasmic rings. In addition, background fluorescence showed the presence of $630 \pm 290$ FliM molecules not associated with spots, and the total number of FliM copies per cell was estimated to be $1,450 \pm 360$. These data yield a fraction from $0.4\times (1450-630)/1450=0.23$ to $0.55\times (1450-630)/1450=0.31$ of FliM copies that may actually be inside full cytoplasmic rings. In other words, $69-77\%$ of FliM copies were outside complete motors. Besides, a fraction from $0.45\times (1450-630)/1450=0.25$ to $0.60\times (1450-630)/1450=0.34$ of FliM copies appeared to belong to partly assembled rings. 
\item In Ref.~\cite{Zhao96} (which studied \textit{S. typhimurium} while the other references studied \textit{E. coli}), whole-cell lysates were separated into three fractions by sedimentation:  $16\% \pm 3$\% of FliM was found to be in the soluble fraction and was thus not part of motors. Gel filtration enabled separation of the larger basal body structures into full flagellar motors and incomplete precursors. Among these large structures, about 31\% of the FliM was found to be part of precursors. At the end of their analysis, the authors also stated that ``about half of FliM remained unaccounted for [in the gel filtration results], suggesting that FliM may form presently uncharacterized, particulate aggregates in addition to being part of flagellar basal bodies". They suggested that those correspond to ``dissociable FliM assembly  intermediates that either get stuck on or elute very late from the column'', thus not appearing in the gel filtration results. Combining this, we may estimate that  a fraction $0.5+0.16+0.31\times(1-0.5-0.16)=0.76$ of FliM copies is not part of complete motors.
\end{itemize}

   \begin{table}[htbp]
  \centering
  \caption{Number of FliM copies per cell and partition.}
    \begin{tabular}{lllllll}
   \hline
  Organism & Total per cell & Outside complete motors & In partly-assembled structures & Ref. \\
    \hline
\textit{E. coli} & 1,400 $\pm$ 200 & 78\% & Not evaluated &\cite{Tang95}\\
\textit{E. coli} & 1,450 $\pm$ 360 & 69-77\%&25-34\% &\cite{Delalez10}\\
\textit{S. typhimurium} & 1,640 $\pm$ 300 & 76\%&31-81\% &\cite{Zhao96}\\
    \hline
    \end{tabular}%
  \label{flim}%
\end{table}

We can also estimate the number of FliM that actually belong to functional flagella. In wild-type \textit{E. coli}, Ref.~\cite{Tang95} reports on average 2.6 flagella per cell, and Ref.~\cite{Turner00} reports about $3\pm1.5$ flagella per cell. In \textit{S. typhimurium}, there are about 6-10 flagella per cell~\cite{Rosu06}. Studies of the stoichiometry of the motor report $37 \pm 13$ FliM copies per motor~\cite{Zhao96, Macnab03} (in \textit{S. typhimurium}), and about 32~\cite{Delalez10} (in \textit{E. coli}). This would give a number of order 100 FliM that actually belong to functional flagella in \textit{E. coli} (about 7\%), and about 300 in \textit{S. typhimurium} (about 18\%). This is even less than what would be expected from the results above (Table~\ref{flim}), especially for \textit{E. coli}. This difference may indicate that some complete or almost complete cytoplasmic rings or motors do not belong to flagella. These might be in the last stages of assembly.

In spite of the high fraction of FliM that are not in functional flagellar motors, the number of flagella per cell increases when FliM is (not too highly) overexpressed, which indicates that FliM constitutes a limiting resource in flagellar assembly~\cite{Tang95}. Besides, underexpression of FliM reduces the number of flagella per cells and their efficiency. Consistently, in Ref.~\cite{Sourjik02}, the number of FliM per cell necessary for optimum motility was about 4900 (about 3.5 times higher than in the wild-type cells), and only about $20-30\%$ of these FliM were found in functional or incomplete flagellar motors (note that these FliM were fluorescently labeled). 

Recent work indicates that the FliM proteins present in motors exchange with the cytoplasmic pool, and that the number of FliM per motor is variable~\cite{Delalez10,Yuan12,Lele12,Yuan13}. The number of FliM copies per motor depends on the concentration of CheY-P~\cite{Yuan12}, through the direction of rotation of the motor~\cite{Lele12}. This allows for adaptation of the motor to the concentration of CheY-P, by shifting the range of CheY-P concentration over which the clockwise bias of the motor changes~\cite{Yuan12}, which is very narrow~\cite{Yuan13}. The fraction of FliM that exchanges depends on the direction of rotation too~\cite{Lele12}. Hence, cytoplasmic FliM seems to have a function in motor adaptation.





\bibliography{ManyChemoreceptors}

\bibliographystyle{biophysj}








\end{document}